\RequirePackage{fix-cm}
\documentclass[smallextended]{svjour3}       
\smartqed  

\def\be{\begin{equation}}
\def\ee{\end{equation}}
\def\bea{\begin{eqnarray}}
\def\eea{\end{eqnarray}}
\def\beaN{\begin{eqnarray*}}
\def\eeaN{\end{eqnarray*}}
\def\ed{\end{document}}
\def\bit{\begin{itemize}}
\def\eit{\end{itemize}}

\def\sig{\sigma}

\def\lam{\lambda}

\def\Del{\Delta}

\def\Bg{\Bar g}

\def\k{\kappa}
\def\alf{\alpha}

\def\BD{\Bar D}

\def\di{\partial}

\def\Lix{\pounds_\xi}

\def\half{{\textstyle{1 \over 2}}}
\def\~{\tilde}
\def\lag{{\hat{\cal L}}}
\def\m{\label}
\def\l{\left}
\def\r{\right}

\def\goto{\rightarrow}
\def\Bar{\overline}
\def\const{\rm const}

\usepackage{graphicx}
\begin{document}

\title{Covariantized N{\oe}ther identities and conservation laws for perturbations in metric theories of gravity
} \subtitle{}

\titlerunning{Conservation laws for perturbations in metric theories of gravity}

\author{Alexander N. Petrov         \and
       \\ Robert R. Lompay 
}


\institute{A. N. Petrov \at
              Moscow MV Lomonosov State University, Sternberg Astronomical
              Institute,  Universitetskii pr. 13, Moscow 119992, Russia \\
              Tel.: +7(495)7315222\\
              Fax: +7(495)9328841\\
              \email{alex.petrov55@gmail.com}           
           \and
                      R. R. Lompay \at
              Department of Physics, Uzhgorod National University,
              Voloshyna str. 54, Uzhgorod 88000, Ukraine\\
              Tel.: +380956940415\\
              \email{rlompay@gmail.com} }

\date{Received: date / Accepted: date}

\maketitle

\begin{abstract}
A construction of conservation laws and conserved quantities for
perturbations in arbitrary metric theories of gravity is developed.
In an arbitrary field theory, with the use of incorporating an
auxiliary metric into the initial Lagrangian covariantized
N{\oe}ther identities are carried out. Identically conserved
currents with corresponding superpotentials are united into a
family. Such a generalized formalism of the covariantized identities
gives a natural basis for constructing conserved quantities for
perturbations. A {\em new family} of conserved currents and
correspondent superpotentials for perturbations on arbitrary curved
backgrounds in metric theories is suggested. The conserved
quantities are both of pure canonical N{\oe}ther and of Belinfante
corrected types. To test the results each of the superpotentials of
the family is applied to calculate the mass of the
Schwarzschild-anti-de Sitter black hole in the Einstein-Gauss-Bonnet
gravity. Using all the superpotentials of the family gives the
standard accepted mass.
 \keywords{Metric gravitation theories \and differential identities\and conservation laws}
\end{abstract}


\section{Introduction}
 \m{Introduction}
 \setcounter{equation}{0}

 Examination and study of perturbations in general relativity (GR),
 including gravitational waves, cosmological perturbations, {\em etc},
 unavoidably lead to constructing  conservation laws and conserved
 quantities for such perturbations.  To describe a perturbed model one needs to choose
 a background spacetime (not perturbed solution of the theory), in which perturbations
 propagate. Depending on a concrete problem, a background can be both flat
 (Minkowski spacetime) and arbitrary curved
 (e.g., cosmological, black hole, or another solution).
 Researches consider such systems both in an approximation (say, in
 linear one) and in an exact form (without any approximation, perturbations
 are not  infinitesimal). In earlier works, conserved quantities in GR, classical
 pseudotensors and superpotentials, have been presented in a non-covariant
 form. At the present time, researches elaborate and study
covariant quantities. For the most of the above one can recommend
the reviews \cite{Petrov2008,Szabados} and numerous references
therein; the paper \cite{PittsSchive2001b} can be very useful also.

As a rule, a construction of conservation laws is based on using the
N{\oe}ther procedure. Even non-covariant pseudotensors and
superpotentials have been connected with the  N{\oe}ther theorem
\cite{Bergmann58,Komar59,Trautman62}. In the framework of GR, the
N{\oe}ther procedure
 is applied to the Einstein-Hilbert action either directly, or not explicitly. Various appropriate
 modifications of the Lagrangian are used, these are an incorporation of background
 structures (such as metric, connection, {\em etc}), an addition of various divergences,
 {\em etc} (see,  e.g., \cite{[11],GPP,KBL,ChenNester,Nester10,Fatibene-etal,PK}). In GR, due to
 efforts of very many authors a technique of applying the N{\oe}ther theorem has
 been well developed. However, last two decades, numerous
 metric theories, which are various modifications of GR, become more
 and more popular. They are quadratic in curvature theories, see, e.g.,
\cite{DT2}; or theories of the Lovelock type \cite{Lovelock}; or
$f(R)$ theories \cite{Sotiriou}, {\em etc}. For them there is also a
necessity to study perturbations and construct conservation laws.
Many results in this direction have been obtained also (see, for
example,
\cite{DerKatzOgushi,Fatibene1,KatzLivshits,Petrov2009,Petrov2009cor,Petrov2009a,Petrov2011},
and reviews \cite{Petrov2008,Szabados} and references therein).
However, concrete forms of metric Lagrangians in such theories are
very various and more complicated than in GR, therefore it is very
desirable to elaborate united rules for constructing conservation
laws for perturbations. The present paper is just devoted to this
problem.

One of the main requirements is to construct covariant expressions
and quantities. However, recall that covariant derivatives of the
metric are equal to zero identically. Therefore a {\em direct }
application of the N{\oe}ther procedure in the framework of
gravitational metric theories leads to {\em non-covariant}
identities and conserved quantities \cite{Mitzk}. To present
covariant expressions one includes either additional structures, or
uses relevant reformulations. One of the earlier attempts to suggest
covariantized N{\oe}ther identities in GR is the Ray work
\cite{Ray}. Here, in a definite sense, we develop Ray's ideas.
Examining perturbed models with a fixed background (a known
solution), we operate with a background metric anyway. Namely its
presence permits us to construct covariant expressions.

Already covariant identities and  conserved quantities for
perturbations in arbitrary field theories with using an auxiliary
background metric have been suggested; these results have been
applied in the framework of the Einstein-Gauss-Bonnet (EGB) gravity
\cite{Petrov2009,Petrov2009cor,Petrov2009a,Petrov2011}. However,
{\em first}, a generalized formalism of constructing such
conservation laws was not presented fully in all the details; {\em
second}, not all the possibilities of such covariant generalized
constructions have been discovered and developed. In the present
paper, we close these gaps. The paper is organized as follows.

In Sect.~\ref{sec:2}, as preliminaries, we give Mitzkevich's
presentation \cite{Mitzk} for deriving N{\oe}ther identities in an
arbitrary field theory. The N{\oe}ther procedure is applied directly
to a {\em covariant} scalar density (Lagrangian), however presented
in a {\em not covariant} form through partial derivatives of field
variables up to a second order. All the metric theories, Lagrangians
of which depend on the Riemann tensor algebraically, are just
related to this class. As a finalized result, not covariant
identically conserved currents and correspondent superpotentials are
obtained.

In Sect.~\ref{sec:3}, {\em covariant} identities and identically
conserved quantities are carried out in an arbitrary field theory.
In Subsect. \ref{sec:31}, for this goal one includes an arbitrary
fixed external metric into the Lagrangian exchanging partial
derivatives (in a special way) by covariant derivatives related to
the external metric. In the result, the initial Lagrangian becomes
evidently covariant, although in a reality it does not contain the
auxiliary metric in whole. This trick permits us to present in a
covariant form both identities and identically conserved quantities.
It turns out that the way of constructing {\em covariant} conserved
quantities presented in Subsect. \ref{sec:31} has a freedom, which
is considered in Subsect. \ref{sec:32}. In Subsect. \ref{sec:33}, we
explore the freedom to present a {\em new family} of covariant
N{\oe}ther identities and identically conserved quantities. In
Subsect. \ref{sec:34}, we expand the {\em new family} using the
Belinfante modification \cite{Belinfante}, which is a more popular
reconstruction of the pure N{\oe}ther procedure \cite{PK}.

In Sect.~\ref{sec:4}, using the generalized results of
Sect.~\ref{sec:3}, we consider an arbitrary {\em metric} theory. In
Subsect.~\ref{sec:41}, necessary elements of the N{\oe}ther
identities related to a {\em pure metric} gravitational Lagrangian
are given. In Subsect.~\ref{sec:42}, a {\em new family} of covariant
conservation laws and conserved quantities for perturbations on
arbitrary curved backgrounds, containing both canonical N{\oe}ther
and Belinfante corrected types, is presented.

In Sect.~\ref{sec:5}, we apply the above results to calculate a mass
of D-dimensional black hole (BH) in the EGB gravity presented by the
Schwartzchild-anti-de Sitter solution \cite{BD+}. In
Subsect.~\ref{sec:51}, we represent the new family of conservation
laws for constructing conserved charges in EGB gravity in general.
In Subsect.~\ref{sec:52}, we use them to calculate the black hole
mass. Already, one of the variants of conservation laws included in
the family has been used to study some solutions in EGB gravity
\cite{Petrov2009,Petrov2009cor,Petrov2009a,Petrov2011}. We take into
account the previous results as a start point for the present
calculations.

In Sect.~\ref{sec:6}, we discuss the obtained results and future
perspectives of their applications.  In Appendix \ref{Appendix1},
useful algebraic formulae for calculating with tensor densities in a
generalized form are presented. In Appendix \ref{Appendix2},
necessary formulae in the EGB gravity are given.

\section{Preliminaries. N{\oe}ther identities in an arbitrary field theory}
 \m{sec:2}

In this section, our presentation follows the presentation in the
book \cite{Mitzk}. We derive the N{\oe}ther identities and conserved
quantities for arbitrary theories, an action of which
 \be
 S= \int  d^Dx\hat L
 \m{actionQ}
 \ee
is invariant with respect to coordinate transformations. Thus, the
action (\ref{actionQ}) is a scalar, whereas the  Lagrangian
 \be
\hat L = \hat L (Q^A; Q^A{}_{,\alf}; Q^A{}_{,\alf\beta}) \m{lagQ}
 \ee
is a scalar density. Here and below `hat' means that a quantity is a
density of the wight +1, for example, $\hat g^{\mu\nu} =
\sqrt{-g}g^{\mu\nu}$, $\hat L = \sqrt{-g}{L}$. Here, the covariance
of the action (or more formally,  of the Lagrangian) is the basis
for applying the N{\oe}ther procedure. Dynamical fields of the
system (\ref{actionQ}) and (\ref{lagQ}) are presented by a set of
tensor densities $Q^A$, where the generalized index `A' is a
collective tensor index. Keeping in mind popular metric
gravitational theories, the Lagrangian includes derivatives up to a
second order. Here and below, Greek indexes numerate coordinates in
$D$-dimensional spacetime; $({,\alf}) \equiv \di_{\alf}$ are partial
(ordinary) derivatives.

Considering Lie displacements as perturbations of the system, we
define variations of fields as their Lie derivatives:
 \be
 \delta_\xi Q^A= {\pounds}_\xi Q^A =-\xi^\alf \di_\alf Q^A +
 {\l. Q^A \r|}^\alf_\beta \di_\alf \xi^\beta\, .
\m{LieQ}
 \ee
Please, note the opposite sign `minus' with respect to the usual one
(we follow the definition of variations in \cite{Mitzk}). The
notation $\l. Q^A \r|^\alf_\beta$ is defined by the transformation
properties of $Q^A$; algebraic properties of $\l. Q^A
\r|^\alf_\beta$,  necessary here, can be found in Appendix
\ref{Appendix1}.

Because the Lagrangian is a scalar density of the weight +1, its
variation leads to the identity:
 \be
\delta_\xi \hat L = {\pounds}_\xi {\hat L} \equiv - (\xi^\alf {\hat
L})_{,\alf} \,.
 \m{LieLag}
 \ee
Keeping in mind a dependence of the Lagrangian (\ref{lagQ}) on $Q^A$
and their derivatives  and substituting (\ref{LieQ}) into
(\ref{LieLag})  one obtains the identity:
 \be
\frac{\delta\hat L}{\delta Q_B}{\pounds}_\xi Q_B + \di_\alf
\l[{{\delta \hat L} \over {\delta Q_{B,\alf}}}{\pounds}_\xi Q_B +
{{\di \hat L} \over {\di Q_{B,\beta\alf}}}({\pounds}_\xi
Q_B)_{,\beta} + \xi^\alf {\hat L}\r]\equiv 0.
 \m{LieIdentity}
 \ee
Usually the identity (\ref{LieIdentity}) is named as the general
N{\oe}ther identity. The Lagrangian derivative is defined as usual
 \be
\frac{\delta\hat L}{\delta Q_B} = \frac{\di\hat L}{\di Q_B} -
\di_\alf \l(\frac{\di\hat L}{\di Q_{B,\alf}}\r)+ \di_{\alf\beta}
\l(\frac{\di\hat L}{\di Q_{B,\alf\beta}}\r)
 \m{L-derivative}
 \ee
(where $\di_{\beta\tau}\equiv \di_\beta\di_\tau$) and defines the
left hand side of the field equations for $Q_B$:
 \be
\frac{\delta\hat L}{\delta Q_B} = 0\,.
 \m{EofM}
 \ee
We introduce also the notation:
 \be
 \m{Lp-derivative}
{{\delta \hat L} \over {\delta  Q_{B,\alf}}}\equiv {{\di \hat L}
\over {\di Q_{B,\alf}}} -
 \di_\beta
\l({{\di \hat L} \over {\di  Q_{B,\alf\beta}}}\r)\,.
 \ee
Substituting (\ref{LieQ}) into (\ref{LieIdentity}) and providing
identical transformations one obtains
 \bea
&-& \l[\frac{\delta\hat L}{\delta Q_B} Q_{B,\alf} + \di_\beta
\l(\frac{\delta\hat L}{\delta Q_B}
\l.Q_{B}\r|^\beta_\alf\r)\r]\xi^\alf\nonumber\\ &+&\di_\alf \l[\hat
U_\sig{}^\alf\xi^\sig + \hat M_{\sig}{}^{\alf\tau}\di_\tau \xi^\sig
+ \hat N_\sig{}^{\alf\tau\beta}\di_{\beta\tau} \xi^\sig\r] \equiv
0\,.
 \m{(+2+)}
 \eea
In (\ref{(+2+)}), the coefficients are defined by the Lagrangian
without ambiguities in unique way:
 \be \hat U_\sig{}^\alf  \equiv  \hat L
\delta^\alf_\sig +
 {{\delta \hat L} \over {\delta Q_B}} \l.Q_B\r|^\alf_\sig -
 {{\delta \hat L} \over {\delta  Q_{B,\alf}}} \di_\sig
Q_{B}  -  {{\di \hat L} \over {\di Q_{B,\beta\alf}}} \di_{\beta\sig}
Q_{B}\, , \m{(+3+)}
 \ee
\be \hat M_\sig{}^{\alf\tau}  \equiv
 {{\delta \hat L} \over {\delta  Q_{B,\alf}}}
 \l.Q_{B}\r|^\tau_\sig -
 {{\di \hat L} \over {\di  Q_{B,\tau\alf}}}
\di_\sig Q_B +
 {{\di \hat L} \over {\di Q_{B,\beta\alf}}}
 \di_\beta (\l.Q_{B}\r|^\tau_\sig)\, ,
\m{(+4+)}
 \ee
\be \hat N_\sig{}^{\alf\tau\beta} \equiv \half \l[{{\di \hat L}
\over {\di  Q_{B,\beta\alf}}}
 \l.Q_{B}\r|^\tau_\sig +
 {{\di \hat L} \over {\di Q_{B,\tau\alf}}}
 \l.Q_{B}\r|^\beta_\sig\r].
\m{(+5+)}
 \ee
To derive the last coefficient the evident relation $\hat
N_\sig{}^{\alf\tau\beta} = \hat N_\sig{}^{\alf\beta\tau}$, following
from (\ref{(+2+)}), has been used.

Opening the identity (\ref{(+2+)}) and, since $\xi^\sig$, $
\di_{\alf}\xi^\sig$, $ \di_{\beta\alf}\xi^\sig$ and
$\di_{\gamma\beta\alf} \xi^\sig$ are arbitrary at every world point,
we equalize to zero the coefficients at them independently and
obtain the system of identities:
 \bea
 &{}& \di_\alf  \hat U_\sig{}^\alf
  \equiv \frac{\delta\hat L}{\delta Q_B} Q_{B,\alf} + \di_\beta
\l(\frac{\delta\hat L}{\delta Q_B}
\l.Q_{B}\r|^\beta_\alf\r), \m{(+9+1A)}\\
&{}&    \hat U_\sig{}^\alf + \di_\lam \hat M_{\sig}{}^{\lam \alf}
\equiv 0,
 \m{(+9+2A)}\\ &{}&
 \hat M_{\sig}{}^{(\alf\beta)}+
\di_\lam  \hat N_{\sig}{}^{\lam(\alf\beta)} \equiv 0, \m{(+9+3A)}\\
&{}&
 \hat
 N^{(\alf\beta\gamma)}_\sig \equiv 0.
 \m{(+9+4A)}
 \eea
As we know, the system of equations (\ref{(+9+1A)}) -
(\ref{(+9+4A)}) was pioneered by Klein \cite{Klein18,Klein21}.
Therefore, we shall refer to this system as {\em the Klein
identities}. After differentiating (\ref{(+9+2A)}) and using
(\ref{(+9+3A)}) and (\ref{(+9+4A)}) one obtains the identity
$\di_\alf  \hat U_\sig{}^\alf \equiv 0$. This means that the right
hand side of (\ref{(+9+1A)}) is equal identically to zero also
 \be
 \frac{\delta\hat L}{\delta Q_B} Q_{B,\alf} + \di_\beta
\l(\frac{\delta\hat L}{\delta Q_B} \l.Q_{B}\r|^\beta_\alf\r) \equiv
0.
 \m{(+9+1A')}
 \ee
These are just the N{\oe}ther identities. Thus instead of the
identity (\ref{(+2+)}) one can use independently (\ref{(+9+1A')})
and
  \be
  \di_\alf \l[\hat
U_\sig{}^\alf\xi^\sig + \hat M_{\sig}{}^{\alf\tau}\di_\tau \xi^\sig
+ \hat N_\sig{}^{\alf\tau\beta}\di_{\beta\tau} \xi^\sig\r] \equiv
0\,.
 \m{(+2+)'}
 \ee
Call the expression under the divergence with a sign `minus' as a
current
 \be
\hat I^\alf(\xi) = -\l[\hat U_\sig{}^\alf\xi^\sig + \hat
M_{\sig}{}^{\alf\tau}\di_\tau \xi^\sig + \hat
N_\sig{}^{\alf\tau\beta}\di_{\beta\tau} \xi^\sig\r]\,.
 \m{currentI}
 \ee
 Sign `minus' is selected to be in a
correspondence with the usual sign `minus' before gravitational
(metric) action  (see, e.g., \cite{Petrov2008}). Because the
divergence (\ref{(+2+)'}) is equal to zero identically, the current
has to be expressed through a quantity (superpotential), a double
divergence of which has to be equal to zero identically: $\hat
I^\alf(\xi) \equiv \di_{\beta}\hat I^{\alf\beta}(\xi)$, where
$\di_{\alf\beta}\hat I^{\alf\beta}(\xi)\equiv 0$. Let us show this.
Due to the symmetry in last two indexes in (\ref{(+5+)}) and the
identity (\ref{(+9+4A)}) one has
 \be
\hat N_\sig{}^{\alf\tau\beta} + \hat N_\sig{}^{\tau\beta\alf} + \hat
N_\sig{}^{\beta\alf\tau} \equiv 0\,.
 \m{N-C-identity}
 \ee
Using this identity and (\ref{(+9+2A)}), after not complicated
transformations one obtains
 \bea
\hat I^\alf(\xi) &=& \di_\beta\l( \hat M_\sig{}^{\beta\alf}\xi^\sig
+ 2\hat N_\sig{}^{\beta\alf\lam}\xi^\sig_{,\lam}  \r)\nonumber\\ &=&
\di_\beta\l[ \hat M_\sig{}^{[\beta\alf]}\xi^\sig + 2\hat
N_\sig{}^{[\beta\alf]\lam}\xi^\sig_{,\lam} - \di_\lam \l(\hat
N_\sig{}^{\lam\alf\beta}\xi^\sig \r)  \r]\,.
 \m{NotCovariantC-S}
 \eea
 This is rewritten in the form of the identity
 \be
 \hat I^\alf(\xi) \equiv \di_\beta  \hat I^{\alf\beta}
 \m{I=IC-S}
 \ee
where the superpotential is defined as
 \be \hat I^{\alf\beta}(\xi) =- \l[ \hat
M_\sig{}^{[\alf\beta]}\xi^\sig + 2\hat
N_\sig{}^{[\alf\beta]\lam}\xi^\sig_{,\lam} + \di_\lam \l(\hat
N_\sig{}^{\lam\alf\beta}\xi^\sig \r)  \r]\,.
 \m{NotCovariantS-l}
 \ee
It is evident that $\di_{\alf\beta}\hat I^{\alf\beta}(\xi)\equiv 0$,
the last term disappears under the double divergence due to the
identity (\ref{(+9+4A)}). Thus (\ref{I=IC-S}) can be considered as
the identity equivalent to the conservation law (\ref{(+2+)'}) for
the current $\hat I^{\alf}(\xi)$.

It is not a goal of the present paper to describe and discuss
nuances of the N{\oe}ther theorems \cite{Kosmann-Schwarzbach} in a
detail. For this important and interesting topic one can see, e.g.,
\cite{KonoplevaPopov,NinaByers,BradingBrown,BradingCastellani,Sardanashvily}
and references there in. Here, the above presentation differs {\em
by the form} from the results of applying the classically formulated
N{\oe}ther theorems. It is because the presentation (\ref{actionQ})
- (\ref{NotCovariantS-l}) is more convenient for deriving conserved
quantities in metric theories of gravity. Although, of course, an
analysis of the general N{\oe}ther identity has to cover the
classical presentation. Let us show this.

The first N{\oe}ther  theorem can be formulated as follows: If the
action $S$ is invariant under a finite continuous group of
transformations $G_r$ depending smoothly on $r$ independent
parameters, then there exist $r$ linearly independent combinations
of the operators of the field equations (Lagrangian derivatives
(\ref{L-derivative})) expressed through divergences of currents. Let
us substitute $\xi^\alf = \varepsilon^\alf = \const$ into
(\ref{LieIdentity}). Then one obtains the identity:
 \be
 \frac{\delta\hat L}{\delta Q_B} Q_{B,\alf} \equiv \di_\beta
\l(\delta_\alf^\beta {\hat L} -{{\delta \hat L} \over {\delta
Q_{B,\beta}}} Q_{B,\alf} - {{\di \hat L} \over {\di
Q_{B,\gamma\beta}}}Q_{B,\alf\gamma} \r) \equiv \di_\beta \hat
J^\beta_{(\alf)}\,,
 \m{LieIdentity+}
 \ee
which is just the result of applying the  first N{\oe}ther theorem.
Here, the currents are defined as $\hat J^\beta_{(\alf)}$ and are
conserved, $\di_\beta\hat J^\beta_{(\alf)}=0$, if the field
equations are satisfied: ${\delta\hat L}/{\delta Q_B}=0$.

The second N{\oe}ther theorem can be formulated as follows: If the
action $S$ is invariant under an infinite continuous group of
transformations $G_{\infty r}$ depending smoothly on $r$ arbitrary
functions, then there exist $r$ identically vanishing linearly
independent combinations of Lagrangian derivatives and their partial
derivatives. The identity (\ref{(+9+1A')}) is the result of applying
this theorem.

Combining the  results of the first and second theorems, N{\oe}ther
has formulated the statement often referred as the third N{\oe}ther
theorem
\cite{KonoplevaPopov,NinaByers,BradingBrown,BradingCastellani,Sardanashvily},
although it was not presented as a theorem.  It is explicitly
presented in section 6: ``An Assertion of Hilbert" of her paper
\cite{Noether1918,Kosmann-Schwarzbach} and sounds as follows: If the
action $S$ is invariant under an infinite continuous group of
transformations $G_{\infty r}$, then the current $\hat
J^\beta_{(\alf)}$ constructed for a finite subgroup $G_r$ of the
group $G_{\infty r}$ (due to the first N{\oe}ther theorem) is the
sum of the divergence of the superpotential and a term vanishing on
the equations of motion. However, N{\oe}ther did not provide a
recipe for the superpotential construction. In the case of generally
covariant theories, an analogous statement was proven by Klein (the
Klein boundary theorem) in his works \cite{Klein18,Klein21} that
appeared prior to the N{\oe}ther's paper where they are cited.
Furthermore, additionally to the identities (\ref{(+9+2A)}) -
(\ref{(+9+4A)}), Klein supplied the recipe for the superpotential
construction. Taking into account such a historical development, we
shall name the system (\ref{(+9+2A)}) - (\ref{(+9+4A)}) and
(\ref{(+9+1A')}) as the {\em Klein-N{\oe}ther identities}.

All the above expressions and identities are not covariant. The next
sections are devoted to a construction of covariant identities and
conserved quantities.

\section{Covariantization of N{\oe}ther identities by including
external (background) metric}
 \m{sec:3}

\subsection{A direct application of the N{\oe}ther procedure}
\m{sec:31}

The Lagrangian (\ref{lagQ}) of a covariant theory is not covariant
evidently. One of the ways to present it in an explicitly covariant
form is to incorporate an external (auxiliary, background) metric as
follows. Turning to the formula (\ref{LieQ+A}) one finds
 \bea
 Q_{B;\alf} &\equiv&  Q_{B,\alf} +  \Bar
\Gamma^\tau_{\alf\rho}\l.Q_B\r|^\rho_\tau\,, \m{Q-part} \\
Q_{B;\alf\beta}& \equiv & Q_{B,\alf\beta} +  \Bar
\Gamma^\tau_{\alf\rho,\beta} \l.Q_B\r|^\rho_\tau + \Bar
\Gamma^\tau_{\alf\rho}\l(\l.Q_B\r|^\rho_\tau\r)_{,\beta}\nonumber\\&+&
\Bar \Gamma^\tau_{\beta\rho}\l[\l.\l(Q_{B}\r|^\rho_\tau\r)_{,\alf} +
\Bar\Gamma^\mu_{\alf\nu}\l. \l.Q_B\r|^\rho_\tau\r|^\nu_\mu -
\delta^\rho_\alf \l(Q_{B,\tau} +
\Bar\Gamma^\mu_{\tau\nu} \l.Q_B\r|^\nu_\mu\r)\r] \nonumber\\
& \equiv & Q_{B,\alf\beta} + \Bar \Gamma^\tau_{\alf\rho,\beta}
\l.Q_B\r|^\rho_\tau + \Bar
\Gamma^\tau_{\alf\rho}\l(\l.Q_B\r|^\rho_\tau\r)_{;\beta} +\Bar
\Gamma^\tau_{\beta\rho}\l.\l(Q_{B;\alf}\r)\r|^\rho_\tau
\nonumber\\&-& \Bar
\Gamma^\tau_{\alf\rho}\Bar\Gamma^\mu_{\beta\nu}\l.
\l.Q_B\r|^\rho_\tau\r|^\nu_\mu \,. \m{Q-2part}
 \eea
Because the Lagrangian (\ref{lagQ}) is a scalar density, after
identical substitutions
 \bea
 Q_{B,\alf}& \equiv & Q_{B;\alf} -  \Bar
\Gamma^\tau_{\alf\rho}\l.Q_B\r|^\rho_\tau\,, \m{Q-part+}\\
Q_{B,\alf\beta}& \equiv & Q_{B;\alf\beta} -  \Bar
\Gamma^\tau_{\alf\rho,\beta} \l.Q_B\r|^\rho_\tau - \Bar
\Gamma^\tau_{\alf\rho}\l(\l.Q_B\r|^\rho_\tau\r)_{;\beta}- \Bar
\Gamma^\tau_{\beta\rho}\l.\l(Q_{B;\alf}\r)\r|^\rho_\tau
\nonumber\\&+& \Bar
\Gamma^\tau_{\alf\rho}\Bar\Gamma^\mu_{\beta\nu}\l.
\l.Q_B\r|^\rho_\tau\r|^\nu_\mu\,, \m{Q-2part+}
 \eea
it is transformed into an explicitly covariant form:
 \be
\hat L(Q_B,\,Q_{B,\alf},\,Q_{B,\alf\beta}) \equiv
\lag(Q_B,\,Q_{B;\alf},\,Q_{B;\alf\beta},\,\Bar g_{\mu\nu},\,\Bar
R^\alf{}_{\mu\beta\nu})\,.
 \m{LagCovariant}
 \ee
Here, $\Bar g_{\mu\nu}$, $\Bar\Gamma^\mu_{\alf\nu}$ and $\Bar
R^\alf{}_{\mu\beta\nu}$ are metric, Cristoffel symbols and the
curvature tensor of the auxiliary spacetime; $({}_{;\alf}) = \Bar
D_\alf$ means a covariant derivative with respect to $\Bar
g_{\mu\nu}$;  here and below `bar' means that a quantity is a
background one. One needs to make an important remark. The left hand
side of (\ref{Q-2part+}) is evidently symmetric in $\alf$ and
$\beta$. To show this for the right hand side one has to present
$Q_{B;\alf\beta}= Q_{B;(\alf\beta)}+Q_{B;[\alf\beta]}$, turn to the
formula (\ref{line-Riemann}) and make necessary algebraic
transformations using other formulae from Appendix \ref{Appendix1}.

To conserve the explicit covariance under variation of $\lag$ the
direct way is to variate the external metric $\Bar g_{\mu\nu}$
together with fields $Q_B$. However, this way is very cumbersome,
and we are going by a more economical one. It is easily to check
that the right hand sides of (\ref{Q-part+}) and (\ref{Q-2part+}) do
not contain the background metric and Christoffel symbols in a
reality. To be convinced in this one has to open covariant
derivatives in the explicit form. Then it is clear that substitution
of (\ref{Q-part+}) and (\ref{Q-2part+}) does not incorporate an
additional external metric. Recall that the initial Lagrangian (the
left hand side of (\ref{LagCovariant})) does not contain a
background metric by definition. Therefore the new presentation of
the initial Lagrangian after substitution of (\ref{Q-part+}) and
(\ref{Q-2part+}), namely  $\lag$ in (\ref{LagCovariant}), does not
contain $\Bar g_{\mu\nu}$ and its derivatives in whole. This means
that finally variation of $\lag$ has to be transformed into the
identity (\ref{(+2+)}) anyway. Thus, we follow the inverse way.
Using (\ref{LagCovariant}), we represent (\ref{(+2+)}) into an
explicitly covariant form, and then obtain covariant identities and
covariant conserved quantities.

At first we note that the identity (\ref{(+2+)}) is covariant in
whole since it has been obtained from the covariant identity
(\ref{LieLag}) directly and conserving all the terms. Now, turn to
the identity (\ref{(+9+1A')}). It is known that the Lagrangian
derivative of the scalar density (\ref{L-derivative}) is covariant.
Here, it is useful to demonstrate this fact and to have a covariant
expression at hand. We use the algebraic properties of the
quantities $\l.Q_B\r|^\rho_\tau$ given in Appendix \ref{Appendix1}.
Let us consider the terms of (\ref{L-derivative}) separately, the
first one can be represented as
 \bea
\frac{\di \hat L}{\di Q_B}& = &\frac{\di \lag}{\di Q_B} + \frac{\di
\lag}{\di Q_{C;\alf}} \frac{\di Q_{C;\alf}}{\di Q_B}+ \frac{\di
\lag}{\di Q_{C;\alf\beta}} \frac{\di Q_{C;\alf\beta}}{\di Q_B}
\nonumber\\&=&\frac{\di \lag}{\di Q_B} + \frac{\di \lag}{\di
Q_{C;\alf}} \frac{\di }{\di
Q_B}\l(\Bar\Gamma^\rho_{\alf\tau}\l.Q_C\r|_\rho^\tau \r)+ \frac{\di
\lag}{\di Q_{C;\alf\beta}} \frac{\di }{\di
Q_B}\l[\Bar\Gamma^\rho_{\alf\tau,\beta}\l.Q_C\r|_\rho^\tau
\r.\nonumber\\&+&\l. \l( \delta^\sig_\alf
\l.\l.Q_C\r|_\rho^\tau{}\r|^\mu_\nu
-\delta^\sig_\rho\delta^\tau_\alf \l.Q_C\r|^\mu_\nu
\r)\Bar\Gamma^\rho_{\beta\tau}\Bar\Gamma^\nu_{\mu\sig}\r]\,.
 \m{firstLd}
 \eea
To derive the second term in (\ref{L-derivative})  it is necessary
the next equality, which follows from (\ref{Q-2part}) and
(\ref{LagCovariant}):
 \bea
\frac{\di \hat L}{\di Q_{B,\alf}}& =& \frac{\di \lag}{\di
Q_{B;\alf}} + \frac{\di \lag}{\di Q_{C;\mu\nu}}\frac{\di
Q_{C;\mu\nu} }{\di Q_{B;\alf}}\nonumber\\& = & \frac{\di \lag}{\di
Q_{B;\alf}} + \frac{\di \lag}{\di Q_{C;\mu\nu}} \frac{\di}{\di
Q_{B;\alf}} \l[\Bar\Gamma^\tau_{\mu\rho}\l(\l.Q_C\r|^\rho_\tau
\r)_{;\nu}+ \Bar\Gamma^\tau_{\nu\rho}\l(\l.Q_{C;\mu}
\r)\r|^\rho_\tau\r].
 \m{1part=cov}
 \eea
Thus for the second term in (\ref{L-derivative}) one has
 \bea \hspace*{-0.5cm}\
-\di_\alf\!\l(\frac{\di \hat L}{\di Q_{B,\alf}}\r)& =& -\bar
D_\alf\! \l(\frac{\di \lag}{\di Q_{B;\alf}}\r) +
\Bar\Gamma^\tau_{\alf\rho}\l.\l(\frac{\di \lag}{\di
Q_{B;\alf}}\r)\r|^\rho_\tau \nonumber\\& - & \di_\alf\!\l( \frac{\di
\lag}{\di Q_{C;\mu\nu}} \frac{\di}{\di Q_{B;\alf}}
\l[\Bar\Gamma^\tau_{\mu\rho}\l(\l.Q_C\r|^\rho_\tau \r)_{;\nu}+
\Bar\Gamma^\tau_{\nu\rho}\l(\l.Q_{C;\mu} \r)\r|^\rho_\tau\r]\r).
 \m{secondLd}
 \eea
To derive the third term in (\ref{L-derivative}) we note that
 \be
 \frac{\di\hat L}{\di Q_{B,\alf\beta}} = \frac{\di\lag}{\di
 Q_{B;\alf\beta}}\,,
 \m{2part=2cov}
 \ee
it is evidently covariant and directly follows from (\ref{Q-2part})
and (\ref{LagCovariant}). (Here, we {\em do not take into account}
the symmetry in $\alf$ and $\beta$ at the left hand side, see
discussion in Subsect.~\ref{sec:32}.) Thus for the third term in
(\ref{L-derivative}) one has
  \bea
\di_{\alf\beta} \l(\frac{\di\hat L}{\di Q_{B,\alf\beta}}\r) &=& \Bar
D_{\alf\beta} \l(\frac{\di\lag}{\di Q_{B;\alf\beta}}\r)-
\Bar\Gamma^\tau_{\alf\rho}\l.\l[\Bar D_\beta \frac{\di\lag}{\di
 Q_{B;\alf\beta}}\r]\r|^\rho_\tau \nonumber\\&-&
\di_\alf \l[\Bar\Gamma^\tau_{\beta\rho}\l.\l( \frac{\di\lag}{\di
 Q_{B;\alf\beta}}\r)\r|^\rho_\tau \r]\,
 \m{thirdLd}
 \eea
where $\BD_{\alf\beta}=\BD_\alf\BD_\beta$. Summing equalities
(\ref{firstLd}), (\ref{secondLd}) and (\ref{thirdLd}) one finds that
the first terms on the right hand sides survive, the other terms are
self-compensated due to the rules in Appendix \ref{Appendix1}. Thus
we show that the Lagrangian derivative (\ref{L-derivative}), the
left hand side of the equations of motion (\ref{EofM}), is
represented in the explicitly covariant form:
 \be
\frac{\delta \hat L}{\delta Q_B} =\frac{\delta \lag}{\delta Q_B} =
\frac{\di \lag}{\di Q_B} -\bar D_\alf \l(\frac{\di \lag}{\di
Q_{B;\alf}}\r)+\Bar D_{\alf\beta} \l(\frac{\di\lag}{\di
Q_{B;\alf\beta}}\r).
 \m{L-derivativeCov}
 \ee
Keeping this in mind and using the properties discussed in Appendix
\ref{Appendix1}, it is not difficult to show that (\ref{(+9+1A')})
has also a covariant form
 \be
 \frac{\delta\lag}{\delta Q_B} Q_{B;\alf} + \Bar D_\beta
\l(\frac{\delta\lag}{\delta Q_B} \l.Q_{B}\r|^\beta_\alf\r) \equiv 0.
 \m{(+9+1A')covar}
 \ee
Then one concludes that the identity (\ref{(+2+)'}), the same as
(\ref{(+2+)}),  is covariant in whole.

Now, let us change partial derivatives of $\xi^\sig$ in
(\ref{(+2+)'}) in the way $\di_\rho\xi^\sig = \Bar D_\rho\xi^\sig -
\l.\xi^\sig\r|^\alf_\beta \Bar \Gamma^\beta_{\rho\alf}$ and rewrite
it as
 \be
 \di_\alf\l[\hat u_\sig{}^\alf\xi^\sig + \hat
m_\sig{}^{\alf\tau}\Bar D_\tau \xi^\sig + \hat
n_\sig{}^{\alf\tau\beta} \Bar D_{\beta\tau} \xi^\sig\r]\equiv 0
 \m{secondIDdi}
 \ee
where
 \bea
 \hat u_\sig{}^\alf &=& \hat U_\sig{}^\alf - \hat
 M_\lam{}^{\alf\tau}\Gamma^\lam_{\sig\tau}+ \hat
 N_\lam{}^{\alf\tau\rho} (\Gamma^\lam_{\tau\pi}\Gamma^\pi_{\sig\rho}
 - \di_\rho\Gamma^\lam_{\sig\tau})\,,\m{Uu}\\
 \hat m_\sig{}^{\alf\tau} &=& \hat M_\sig{}^{\alf\tau} +
 \hat N_\sig{}^{\alf\lam\rho}\Gamma^\tau_{\lam\rho} -
 2\hat N_\lam{}^{\alf\tau\rho}\Gamma^\lam_{\sig\rho}\,,
 \m{Mm}\\
 \hat n_\sig{}^{\alf\tau\beta} &=& \hat N_\sig{}^{\alf\tau\beta}\,.
 \m{Nn}
 \eea
Below, with using the new form of the Lagrangian
(\ref{LagCovariant}) and the connections between partial and
covariant derivatives (\ref{Q-part}) and (\ref{Q-2part}) we show
that the new coefficients (\ref{Uu}) - (\ref{Nn}) are represented in
an explicitly covariant form.

Let us begin from the last the coefficient (\ref{Nn}). Due to
(\ref{2part=2cov}) the coefficient $N$ (\ref{(+5+)}) is
automatically covariant, and the coefficient $n$ is presented in the
obviously covariant form \be
 \hat n_\sig{}^{\alf\tau\beta} \equiv\half \l[{{\di \lag} \over
{\di Q_{B;\beta\alf}}}
 \l.Q_{B}\r|^\tau_\sig +
 {{\di \lag} \over {\di Q_{B;\tau\alf}}}
 \l.Q_{B}\r|^\beta_\sig\r].
\m{(+5+)+}
 \ee
To represent $m$ in (\ref{Mm}) we need in the representation of $M$
in (\ref{(+4+)}). The first term in $M$ is defined by the derivative
(\ref{Lp-derivative}), let us reproduce it. For this we use
(\ref{1part=cov}), (\ref{2part=2cov}), and rules in Appendix
\ref{Appendix1}. One gets
 \bea
 {{\delta \hat L} \over {\delta  Q_{B,\alf}}}
 & = &\frac{\di \lag}{\di
Q_{B;\alf}} + \frac{\di \lag}{\di Q_{B;\mu\nu}} \frac{\di}{\di
Q_{B;\alf}} \l[\Bar\Gamma^\tau_{\mu\rho}\l(\l.Q_B\r|^\rho_\tau
\r)_{;\nu}+ \Bar\Gamma^\tau_{\nu\rho}\l(\l.Q_{B;\mu}
\r)\r|^\rho_\tau\r]\nonumber\\&-& \Bar D_{\beta}
\l(\frac{\di\lag}{\di Q_{B;\alf\beta}}\r)+ \Bar\Gamma^\nu_{\beta\mu}
\l.\l(\frac{\di\lag}{\di Q_{B;\alf\beta}}\r)\r|^\mu_\nu \,.
 \m{Lp-derivative+}
 \eea
Using this expression and (\ref{2part=2cov}), and again rules in
Appendix \ref{Appendix1}, one obtains finally for $M$ in
(\ref{(+4+)}):
 \bea \hat M_\sig{}^{\alf\tau} & \equiv &
\l[\frac{\di \lag}{\di Q_{B;\alf}} + \frac{\di \lag}{\di
Q_{B;\mu\nu}} \frac{\di}{\di Q_{B;\alf}}
\l[\Bar\Gamma^\tau_{\mu\rho}\l(\l.Q_B\r|^\rho_\tau \r)_{;\nu}+
\Bar\Gamma^\tau_{\nu\rho}\l(\l.Q_{B;\mu}
\r)\r|^\rho_\tau\r]\r.\nonumber\\&-& \l.\Bar D_{\beta}
\l(\frac{\di\lag}{\di Q_{B;\alf\beta}}\r)+ \Bar\Gamma^\nu_{\beta\mu}
\l.\l(\frac{\di\lag}{\di Q_{B;\alf\beta}}\r)\r|^\mu_\nu \r]
\l.Q_{B}\r|^\tau_\sig \nonumber\\& - & {{\di \lag} \over {\di
Q_{B;\tau\alf}}} \l[\Bar D_\sig Q_B -
\Bar\Gamma^\nu_{\sig\mu}\l.Q_B\r|^\mu_\nu \r] \nonumber\\&+&
 {{\di \lag} \over {\di Q_{B;\beta\alf}}}
\l[ \Bar D_\beta (\l.Q_{B}\r|^\tau_\sig) -
\Bar\Gamma^\nu_{\beta\mu}\l.\l.Q_B\r|^\tau_\sig\r|^\mu_\nu \r]\, .
\m{(+4+)+}
 \eea
Next, substituting (\ref{(+4+)+}),  (\ref{Nn}) and (\ref{(+5+)+})
into (\ref{Mm}) and, after using the rules of Appendix
\ref{Appendix1}, one gets the evidently covariant form for $m$:
  \bea \hat m_\sig{}^{\alf\tau}  &\equiv&
\l[\frac{\di \lag}{\di Q_{B;\alf}} - \Bar D_{\beta}
\l(\frac{\di\lag}{\di Q_{B;\alf\beta}}\r)\r] \l.Q_{B}\r|^\tau_\sig
\nonumber\\&-& {{\di \lag} \over {\di Q_{B;\tau\alf}}} \Bar D_\sig
Q_B +
 {{\di \lag} \over {\di Q_{B;\beta\alf}}}
\Bar D_\beta (\l.Q_{B}\r|^\tau_\sig) \, . \m{m(+4+)}
 \eea

To derive $u$ in (\ref{Uu}) we have already (\ref{Nn}) with
(\ref{(+5+)+}), and (\ref{(+4+)+}). We need to represent only
(\ref{(+3+)}), where the first two terms are evidently covariant
(see (\ref{LagCovariant}) and (\ref{L-derivativeCov})); the third
term is defined by (\ref{Lp-derivative+}) and by (\ref{Q-part+}):
 \be
- {{\delta \hat L} \over {\delta  Q_{B,\alf}}} \di_\sig Q_{B} = -
{{\delta \hat L} \over {\delta  Q_{B,\alf}}} \l(\Bar D_\sig Q_{B} -
\Bar\Gamma^\tau_{\rho\sig}\l.Q_B\r|^\rho_\tau\r)\,,
 \m{thirdU}
 \ee
and the fourth term is defined by (\ref{2part=2cov}) and
(\ref{Q-2part+})
 \bea  -  {{\di \hat L} \over {\di Q_{B,\beta\alf}}} \di_{\beta\sig}
Q_{B} =&-&  {{\di \lag} \over {\di Q_{B;\beta\alf}}} \l(\Bar
D_{\beta\sig}Q_B -  \Bar \Gamma^\tau_{\sig\rho,\beta}
\l.Q_B\r|^\rho_\tau - \Bar
\Gamma^\tau_{\sig\rho}\l. \Bar D_{\beta} (\l.Q_B\r|^\rho_\tau\r)\r.\nonumber\\
&-& \l.\Bar \Gamma^\tau_{\beta\rho}\l.\l(Q_{B;\sig}\r)\r|^\rho_\tau
+ \Bar \Gamma^\tau_{\sig\rho}\Bar\Gamma^\mu_{\beta\nu}\l.
\l.Q_B\r|^\rho_\tau\r|^\nu_\mu \r).
 \m{fourthU}
 \eea
Finally one has the evidently covariant form:
 \bea \hat u_\sig{}^\alf & \equiv & \lag
\delta^\alf_\sig +
 \frac{\delta \lag}{\delta Q_B}
\l.Q_B\r|^\alf_\sig -\l[\frac{\di \lag}{\di Q_{B;\alf}} - \Bar
D_{\beta} \l(\frac{\di\lag}{\di Q_{B;\alf\beta}}\r)\r] \Bar D_\sig
Q_{B}  \nonumber\\ & -&  {{\di \lag} \over {\di Q_{B;\beta\alf}}}
\Bar D_{\beta\sig} Q_{B}+ \frac{1}{2} {{\di \lag} \over {\di
Q_{B;\tau\alf}}}\l.Q_B\r|^\beta_\lam \Bar R^\lam{}_{\sig\tau\beta}\,
. \m{u(+3+)}
 \eea

Showing that the coefficients in (\ref{secondIDdi}) (rewritten
(\ref{(+2+)'})) are covariant, we demonstrate that the expression
under divergence in (\ref{secondIDdi}) in whole is a vector density,
and  the identity (\ref{secondIDdi}) can be rewritten as
 \be
 \BD_\alf\l[\hat u_\sig{}^\alf\xi^\sig + \hat
m_\sig{}^{\alf\tau}\Bar D_\tau \xi^\sig + \hat
n_\sig{}^{\alf\tau\beta} \Bar D_{\beta\tau} \xi^\sig\r]\equiv 0.
 \m{secondID}
 \ee
Opening it and equating independently to zero the coefficients at
$\xi^\sig$, $\BD_{\alf} \xi^\sig$, $\BD_{(\beta\alf)} \xi^\sig$ and
$\BD_{(\gamma\beta\alf)} \xi^\sig$, we get a set of identities:
 \bea
 &{}& \BD_\alf  \hat u_\sig{}^\alf + \half
 \hat m_\lam{}^{\alf\rho} \Bar R^{~\lam}_{\sig~\rho\alf}
 +{\textstyle{1\over 3}} \hat n_\lam{}^{\alf\rho\gamma}
\BD_\gamma \Bar R^{~\lam}_{\sig~\rho\alf}
  \equiv 0, \m{(+9+1)}\\
&{}&    \hat u_\sig{}^\alf + \BD_\lam \hat m_{\sig}{}^{\lam \alf} +
\hat n_\lam{}^{\tau\alf\rho}
 \Bar R^{~\lam}_{\sig~\rho\tau} +{\textstyle{2\over 3}} \hat
 n_{\sig}{}^{\lam\tau\rho}\Bar R^{\alf}_{~\tau\rho\lam} \equiv 0,
 \m{(+9+2)}\\ &{}&
 \hat m_{\sig}{}^{(\alf\beta)}+
\BD_\lam  \hat n_{\sig}{}^{\lam(\alf\beta)} \equiv 0, \m{(+9+3)}\\
&{}&
 \hat
 n^{(\alf\beta\gamma)}_\sig \equiv 0.
 \m{(+9+4)}
 \eea
Substituting here the initial definitions (\ref{Uu}) - (\ref{Nn})
one can be convinced that the system (\ref{(+9+2)}) - (\ref{(+9+4)})
consists of linear combinations of the Klein identities
(\ref{(+9+2A)}) - (\ref{(+9+4A)}). The identity (\ref{(+9+1)})
corresponds to $\di_\alf \hat U^\alf_\sig \equiv 0$. The last is a
consequence of (\ref{(+9+2A)}) - (\ref{(+9+4A)}). Analogously,
(\ref{(+9+1)}) is not independent - it is a consequence of
(\ref{(+9+2)}) - (\ref{(+9+4)}).

Since the equality (\ref{secondID}) is identically satisfied, the
current
 \be
 \hat \imath^\alf =-\l[\hat u_\sig{}^\alf\xi^\sig + \hat
m_\sig{}^{\alf\tau}\Bar D_\tau \xi^\sig + \hat
n_\sig{}^{\alf\tau\beta} \Bar D_{\beta\tau} \xi^\sig\r] \,,
 \m{(+7+)}
 \ee
 must be a divergence of a superpotential (antisymmetrical tensor density):
 \be  \hat
\imath^{\alf}\equiv \di_\beta \hat \imath^{\alf\beta},
 \m{(+10+)}
 \ee
for which $\di_{\beta\alf} \hat \imath^{\alf\beta} \equiv 0$.
Indeed, substituting $\hat u_\sig{}^\alf$  from (\ref{(+9+2)})
 into the current (\ref{(+7+)}), using (\ref{(+9+3)}) and algebraic properties of
$\hat n_\sig{}^{\alf\beta\gamma}$ and $\Bar
R^{\alf}_{~\beta\rho\sig}$, and conserving the covariance, we
reconstruct (\ref{(+7+)}) into the form (\ref{(+10+)}), where  the
superpotential is
 \be
 \hat \imath^{\alf\beta}  = \l({\textstyle{2\over 3}}
 \BD_\lam  \hat n_{\sig}{}^{[\alf\beta]\lam}  - \hat
 m_{\sig}{}^{[\alf\beta]}\r)\xi^\sig   -
 {\textstyle{4\over 3}} \hat n_{\sig}{}^{[\alf\beta]\lam}
 \BD_\lam  \xi^\sig.
 \m {(+11+)}
 \ee
It is explicitly antisymmetric in $\alf$ and $\beta$.

At last, the current in (\ref{(+7+)}) can be rewritten as
 \be
 \hat \imath^\alf =- \l[(\hat u_\sig{}^\alf + \hat n_\lam{}^{\alf\beta\gamma}
\Bar R^\lam_{~\beta\gamma\sig})\xi^\sig + \hat
 m^{\rho\alf\beta}\di_{[\beta}\xi_{\rho]} +
 \hat z^{\alf}\r]
 \m{(+7+A)}
 \ee
where $z$-term  is defined as
 \be
 \hat z^{\alf} (\xi) =\hat
m^{\sig\alf\beta}\zeta_{\sig\beta}+ \hat n^{\rho\alf\beta\gamma}
\l(2 \BD_{\gamma}\zeta_{\beta\rho} - \BD_\rho
\zeta_{\beta\gamma}\r)\,,
 \m{(+8+)}
 \ee
and $2\zeta_{\rho\sigma} = - {\pounds}_\xi \Bg_{\rho\sigma} =
2\BD_{(\rho}\xi_{\sigma)}$. Thus, $z$-term disappears, if $\xi^\mu $
is a Killing vector of the background spacetime. Then only the
current (\ref{(+7+A)}) is determined by the energy-momentum $(u +
n\Bar R)$-term and the spin $m$-term.

Now, let us sum the results. Instead of the non-covariant
coefficients (\ref{(+3+)}), (\ref{(+4+)}) and (\ref{(+5+)}),
correspondent covariant coefficients (\ref{(+5+)+}), (\ref{m(+4+)})
and (\ref{u(+3+)}) have been constructed. Instead of the
non-covariant Klein-N{\oe}ther identities (\ref{(+9+2A)}) -
(\ref{(+9+4A)}), (\ref{(+9+1A')}),  correspondent covariant
identities (\ref{(+9+2)}) - (\ref{(+9+4)}), (\ref{(+9+1A')covar})
are presented. By the construction, the explicitly covariant current
(\ref{(+7+)}) is equal to the current (\ref{NotCovariantC-S}) in the
original form exactly: $ \hat \imath^{\alf}\equiv \hat I^{\alf}$.
However, one can show that
 \be
 \hat\imath^{\alf\beta} = \hat I^{\alf\beta} +{\textstyle \frac{4}{3}}\di_\lam
 \l(\hat N_\sig{}^{[\lam\beta]\alf}\xi^\sig \r)\,.
 \m{iab=Iab}
 \ee
 This means, of course, that $ \di_\beta \hat
\imath^{\alf\beta}\equiv \di_\beta \hat I^{\alf\beta}$ for the
superpotentials (\ref{(+11+)}) and (\ref{NotCovariantS-l}), thus
there is no a contradiction.

\subsection{Another variant of covariantization }
 \m{sec:32}

In Sect.~\ref{sec:2}, all the identities and conserved quantities
are derived through partial derivatives. The order of partial
derivatives is not important because they are symmetrical with
respect to replacements. For example, expressions, like ${{\di \hat
L} / {\di Q_{B,\alf\beta}}}$, are symmetrical in $\alf$ and $\beta$.
Nevertheless,  in previous subsection, we {\em did not used} the
symmetry of partial derivatives, {\em conserving an original order}
of derivatives in the identities (unlike \cite{Mitzk}), see remark
after (\ref{2part=2cov})). This has permitted us to present the
covariant versions of identities and conserved quantities. However,
there are another possibilities.

To introduce the situation let us consider an auxiliary Lagrangian
$\hat L^{test}=\hat P^{B\alf\beta}Q_{B,\alf\beta}+ \ldots $ as an
example. After direct covariantization it acquires the form
$\lag^{test}=\hat P^{B\alf\beta}Q_{B;\alf\beta}+ \ldots $. The
variation with respect to $Q_{B,\alf\beta}$ in the fist case gives
$P^{B(\alf\beta)}$. However, originally $P^{B\alf\beta}$ is not
necessarily symmetrical in $\alf$ and $\beta$, therefore in the
second case the variation with respect to $Q_{B;\alf\beta}$ gives
simply $P^{B\alf\beta}$. Thus, unlike the first case, the other
order of second covariant derivatives can lead to a different
result. If we symmetrize $\alf$ and $\beta$ in the second case:
$\lag^{test}=\hat P^{B\alf\beta}Q_{B;(\alf\beta)}+ \ldots $ then we
need to change the other terms in the Lagrangian.

To study the problem of a different order of second covariant
derivatives let us change this order in (\ref{LagCovariant}):
   \bea
\hat L(Q_B,\,Q_{B,\alf},\,Q_{B,\alf\beta}) &\equiv &
\lag(Q_B,\,Q_{B;\alf},\,Q_{B;\alf\beta},\,\Bar g_{\mu\nu},\,\Bar
R^\alf{}_{\mu\beta\nu})\nonumber \\&\equiv &
\lag(Q_B,\,Q_{B;\alf},\,Q_{B;\beta\alf}+\l.Q_B\r|^\rho_\sig\Bar
R_\rho{}^\sig{}_{\alf\beta},\,\Bar g_{\mu\nu},\,\Bar
R^\alf{}_{\mu\beta\nu})\nonumber \\&\equiv &
\lag^*(Q_B,\,Q_{B;\alf},\,Q_{B;\alf\beta},\,\Bar g_{\mu\nu},\,\Bar
R^\alf{}_{\mu\beta\nu})\,.
 \m{LagCovariant+}
 \eea
For the sake of clearance one has to explain the notations. Here,
the second line has the {\em form} of the Lagrangian of the first
line, only the inverse second covariant derivatives are used.
One can see that the arguments are mixed at the second line. After
re-ordering the arguments, following the first line, it is clear
that the Lagrangian acquires the {\em other form} (the third line).
The star form is useful for the presentation because we need not
remark every time that we use the inverse order of derivatives.

After the exchange in (\ref{LagCovariant+}), it is evidently that
derivatives with respect to second covariant derivatives of $Q_B$
change their order, also one obtains an additional derivative with
respect to $Q_B$, proportional to the Riemannian tensor.  At first,
one has to be convinced that after this exchange the equations of
motion (\ref{EofM}) do not change. Of course, for the starred
Lagrangian the form of the Lagrangian derivative has to be the same
(\ref{L-derivativeCov}). Then, substituting the second line of
(\ref{LagCovariant+}) into (\ref{L-derivativeCov}) one obtains
 \be
\frac{\delta \lag^*}{\delta Q_B}=\frac{\di \lag}{\di Q_B} -\bar
D_\tau \l(\frac{\di \lag}{\di Q_{B;\tau}}\r)+\Bar D_{\tau\beta}
\l(\frac{\di\lag}{\di Q_{B;\beta\tau}}\r) +\frac{\di\lag}{\di
Q_{C;\tau\beta}}\frac{\di}{\di Q_B}\l(\l.Q_C\r|^\rho_\lam\Bar
R_\rho{}^\lam{}_{\tau\beta} \r).
 \m{EofM+}
 \ee
Changing the order of derivatives in the third term at the right
hand side, using (\ref{line-Riemann}) and other formulae in Appendix
\ref{Appendix1}, one can see that, indeed, it is same Lagrangian
derivative (\ref{L-derivativeCov}). However, the different
definitions of the covariantized Lagrangin in (\ref{LagCovariant+})
lead to different conserved quantities that we show below.

The use of the second line of (\ref{LagCovariant+}) in
(\ref{(+5+)+}), (\ref{m(+4+)}) and (\ref{u(+3+)}) gives
 \bea
 \hat n^*_\sig{}^{\alf\tau\beta} &\equiv &\half \l[{{\di \lag} \over
{\di Q_{B;\alf\beta}}}
 \l.Q_{B}\r|^\tau_\sig +
 {{\di \lag} \over {\di Q_{B;\alf\tau}}}
 \l.Q_{B}\r|^\beta_\sig\r],
\m{n-inverse}\\
\hat m^*_\sig{}^{\alf\tau} & \equiv &\l[\frac{\di \lag}{\di
Q_{B;\alf}} - \Bar D_{\beta} \l(\frac{\di\lag}{\di
Q_{B;\beta\alf}}\r)\r] \l.Q_{B}\r|^\tau_\sig - {{\di \lag} \over
{\di Q_{B;\alf\tau}}} \Bar D_\sig Q_B \nonumber\\&+&
 {{\di \lag} \over {\di Q_{B;\alf\beta}}}
\Bar D_\beta (\l.Q_{B}\r|^\tau_\sig), \m{m-inverse}
 \\
  \hat u^*_\sig{}^\alf & \equiv & \lag
\delta^\alf_\sig +
 \frac{\delta \lag}{\delta Q_B} \l.Q_B\r|^\alf_\sig -
 \l[\frac{\di \lag}{\di Q_{B;\alf}} - \Bar D_{\beta} \l(\frac{\di\lag}{\di Q_{B;\beta\alf}}\r)\r] \Bar D_\sig
Q_{B} \nonumber\\& -& {{\di \lag} \over {\di Q_{B;\alf\beta}}} \Bar
D_{\beta\sig} Q_{B}+ \frac{1}{2} {{\di \lag} \over {\di
Q_{B;\alf\tau}}}\l.Q_B\r|^\beta_\lam \Bar R^\lam{}_{\sig\tau\beta}\,
. \m{u-inverse}
 \eea
Remark, to obtain the set (\ref{n-inverse}) - (\ref{u-inverse}) one
has to turn to the expressions (\ref{(+5+)+}), (\ref{m(+4+)}) and
(\ref{u(+3+)}) and change the order of the second derivatives in all
the terms, like ${{\di \lag} / {\di Q_{B;\alf\beta}}} \goto {{\di
\lag} / {\di Q_{B;\beta\alf}}}$, {\em simultaneously}.

Because all the expressions (\ref{n-inverse}) - (\ref{u-inverse})
have been obtained from the Lagrangian $\lag^*$ that is a scalar
density, like  $\lag$, they have to satisfy all the same identities
(\ref{(+9+1)}) - (\ref{(+9+4)}) also. Let us show this. It is not
difficult to find a connection of the expressions (\ref{n-inverse})
- (\ref{u-inverse}) with the coefficients (\ref{(+5+)+}),
(\ref{m(+4+)}) and (\ref{u(+3+)}): \bea
 \hat n^*_\sig{}^{\alf\tau\beta} &\equiv &
 \hat n_\sig{}^{\alf\tau\beta}+{{\di \lag} \over
{\di Q_{B;[\alf\beta]}}}
 \l.Q_{B}\r|^\tau_\sig +
 {{\di \lag} \over {\di Q_{B;[\alf\tau]}}}
 \l.Q_{B}\r|^\beta_\sig.
\m{n-inverse+}\\
\hat m^*_\sig{}^{\alf\tau} & \equiv & \hat m_\sig{}^{\alf\tau}
 - 2{{\di \lag} \over {\di Q_{B;[\alf\tau]}}} \Bar D_\sig Q_B +
 2\Bar D_\beta\l({{\di \lag} \over {\di Q_{B;[\alf\beta]}}}
 \l.Q_{B}\r|^\tau_\sig\r) \, . \m{m-inverse+}
 \\
  \hat u^*_\sig{}^\alf & \equiv & \hat u_\sig{}^\alf   - 2
\Bar D_{\beta}\l({{\di \lag} \over {\di Q_{B;[\alf\beta]}}} \Bar
D_{\sig} Q_{B}\r)+ {{\di \lag} \over {\di
Q_{B;[\alf\tau]}}}\l.Q_B\r|^\beta_\lam \Bar
R^\lam{}_{\sig\tau\beta}\,  \m{u-inverse+}
 \eea
where we define
 \be
 {{\di \lag} \over {\di
Q_{B;[\alf\beta]}}} \equiv \frac12 \l( {{\di \lag} \over {\di
Q_{B;\alf\beta}}} - {{\di \lag} \over {\di Q_{B;\beta\alf}}}\r). \ee
A direct substitution of (\ref{n-inverse+}) - (\ref{u-inverse+})
into the identities (\ref{(+9+1)}) - (\ref{(+9+4)}) shows that
$n^*$, $m^*$ and $u^*$ satisfy them also, like the coefficients $n$,
$m$ and $u$. This means that if we construct a starred current with
using the rule (\ref{(+7+)}):
  \be
 \hat \imath^{*\alf} =-\l[\hat u^*_\sig{}^\alf\xi^\sig + \hat
m^*_\sig{}^{\alf\tau}\Bar D_\tau \xi^\sig + \hat
n^*_\sig{}^{\alf\tau\beta} \Bar D_{\beta\tau} \xi^\sig\r] \,,
 \m{(+7+star)}
 \ee
then it is conserved identically. Indeed, it is easily  to find that
 \be
 \hat \imath^{*\alf} = \hat \imath^\alf -
 2\Bar D_\beta\l({{\di \lag} \over {\di Q_{B;[\alf\beta]}}}
\pounds_\xi Q_{B}\r).
 \m{Cimath+}
 \ee
Then  $\di_\alf \hat \imath^{*\alf}\equiv \di_\alf \hat
\imath^{\alf}$, and consequently  $\di_\alf \hat
\imath^{*\alf}\equiv 0$.  Analogously to (\ref{(+10+)}), the
identity
 \be  \hat
\imath^{*\alf}\equiv \di_\beta \hat \imath^{*\alf\beta}
 \m{(+10+star)}
 \ee
exists where
 \be
 \hat \imath^{*\alf\beta}  = \l({\textstyle{2\over 3}}
 \BD_\lam  \hat n^*_{\sig}{}^{[\alf\beta]\lam}  - \hat
 m^*_{\sig}{}^{[\alf\beta]}\r)\xi^\sig   -
 {\textstyle{4\over 3}} \hat n^*_{\sig}{}^{[\alf\beta]\lam}
 \BD_\lam  \xi^\sig\,.
 \m {(+11+star)}
 \ee
 The direct substitution  of (\ref{n-inverse+}) and (\ref{m-inverse+}) into (\ref{(+11+star)}) gives
  \bea \hspace*{-0.5cm}
 \hat \imath^{*\alf\beta} &=& \hat \imath^{\alf\beta} -
2{{\di \lag} \over {\di Q_{B;[\alf\beta]}}} \pounds_\xi
Q_{B}\nonumber\\& +& \frac{2}{3}\BD_\rho\l[\xi^\sig\!\!\l({{\di
\lag} \over {\di Q_{B;[\alf\beta]}}}
 \l.Q_{B}\r|^\rho_\sig - {{\di \lag} \over
{\di Q_{B;[\alf\rho]}}}
 \l.Q_{B}\r|^\beta_\sig + {{\di \lag} \over
{\di Q_{B;[\beta\rho]}}}
 \l.Q_{B}\r|^\alf_\sig \r)\r]\!.
 \m{Simath+A}
 \eea
The expression in the square brackets is antisymmetric in $\alf$,
$\beta$ and $\rho$. Thus $\BD_\rho$ can be changed by $\di_\rho$ and
one can see that the term in the square brackets does not contribute
into the current in (\ref{(+10+star)}). Also, due to the Stockes
theorem this term does not contribute into surface integrals
calculated with the use of the superpotential. Therefore we use only
 \be
 \hat \imath^{*\alf\beta} = \hat \imath^{\alf\beta} -
 2{{\di \lag} \over {\di Q_{B;[\alf\beta]}}}
\pounds_\xi Q_{B},
 \m{Simath+}
 \ee
which is in a correspondence with (\ref{Cimath+}) and
(\ref{(+10+star)}).

\subsection{A new family of the covariant N{\oe}ther identically
conserved quantities} \m{sec:33}

It is worthy to discuss the situation. One has the identities
 \be \di_\alf \hat I^\alf \equiv
\di_\alf \hat \imath^\alf \equiv \di_\alf \hat \imath^{*\alf}\equiv
0.
 \m{identitiesStar+}
 \ee
Recall that each of the Lagrangains $\hat L$, $\lag$ and $\lag^*$
gives the same equations of motion for $Q_B$. Then adding
(\ref{(+9+1A')}) in a related form to each of the identities in
(\ref{identitiesStar+}) one obtains
  \bea
 -\frac{\delta\hat L}{\delta Q_B} Q_{B,\alf} - \di_\beta
\l(\frac{\delta\hat L}{\delta Q_B} \l.Q_{B}\r|^\beta_\alf\r) -
\di_\alf \hat I^\alf  & \equiv & 0\,,
 \m{I+9}\\
 -\frac{\delta\lag}{\delta Q_B} Q_{B;\alf} - \Bar D_\beta
\l(\frac{\delta\lag}{\delta Q_B} \l.Q_{B}\r|^\beta_\alf\r) -
\di_\alf \hat \imath^\alf   & \equiv & 0\,,
 \m{i+9}\\
 -\frac{\delta\lag^*}{\delta Q_B} Q_{B;\alf} - \Bar D_\beta
\l(\frac{\delta\lag^*}{\delta Q_B} \l.Q_{B}\r|^\beta_\alf\r) -
\di_\alf \hat \imath^{*\alf}   & \equiv & 0\,.
 \m{i+9star}
 \eea
Each of these identities is its own form of the unique identity
(\ref{(+2+)}). All of the identities (\ref{I+9}) - (\ref{i+9star})
can be interpreted as following after variation of the same
Lagrangian. Therefore, the choice of both $\hat \imath^{\alf}$ and
$\hat \imath^{*\alf}$, as a covariantized current, has equal rights.
Nevertheless, what could be preferable from them? From the first
glance it seems that it is $\hat \imath^{\alf}$ because by the
construction, $\hat \imath^{\alf} \equiv \hat I^\alf $. On the other
hand, a conservation of the symmetry of partial derivatives looks as
a nice idea. Then one can choose $\hat
L(Q_B,\,Q_{B,\alf},\,Q_{B,\alf\beta}) \equiv
\lag(Q_B,\,Q_{B;\alf},\,Q_{B;(\alf\beta)}+\half\l.Q_B\r|^\rho_\sig\Bar
R_\rho{}^\sig{}_{\alf\beta},\,\Bar g_{\mu\nu},\,\Bar
R^\alf{}_{\mu\beta\nu})$ instead of (\ref{LagCovariant+}). In a
reality, we do not see any theoretical foundation for a choice.
Possibly, in future, applications to complicated solutions of the
numerous modern modifications of GR will permit to do the choice. To
unite aforementioned possibilities for constructing covariant
conserved quantities we suggest a covariantized Lagrangian of the
form:
 \bea
\hat L(Q_B,\,Q_{B,\alf},\,Q_{B,\alf\beta})&\equiv&
\lag^\dagger(Q_B,\,Q_{B;\alf},\,Q_{B;\alf\beta},\,\Bar
g_{\mu\nu},\,\Bar
R^\alf{}_{\mu\beta\nu})\nonumber\\
&\equiv & p\,\lag(Q_B,\,Q_{B;\alf},\,Q_{B;\alf\beta},\,\Bar
g_{\mu\nu},\,\Bar
R^\alf{}_{\mu\beta\nu})\nonumber\\&+&q\,\lag^*(Q_B,\,Q_{B;\alf},\,Q_{B;\alf\beta},\,\Bar
g_{\mu\nu},\,\Bar R^\alf{}_{\mu\beta\nu})
 \m{p+q}
 \eea
where $p+q=1$ with real $p$ and $q$. The Lagrangian (\ref{p+q})
leads to the same field equations (\ref{EofM}), whereas the
conservation law and conserved quantities for (\ref{p+q}) are
defined now as
 \bea
\hat\imath^{\dagger\alf}&\equiv& \di_\beta \hat
\imath^{\dagger\alf\beta}\,,
 \m{(+10+dagger)} \\
 \hat \imath^{\dagger\alf} &\equiv& p\,\hat\imath^{\alf} + q\,\hat\imath^{*\alf}\,,
 \m{Cdagger}\\
 \hat\imath^{\dagger\alf\beta} &\equiv& p\,\hat\imath^{\alf\beta} +
 q\,\hat\imath^{*\alf\beta}\,,
 \m{Sdagger}
 \eea
presenting in a reality a family of identically conserved
quantities.

To finalize subsection one has to note the following. Recall that a
divergence in the Lagrangian, being non-essential for deriving field
equations, is important (frequently even crucial)  in a definition
of N{\oe}ther canonical conserved quantities. We give some necessary
formulae. For the scalar density $\lag'= \hat d^\nu{}_{,\nu}$ one
has the N{\oe}ther identity $ ({\Lix} \hat d^\alf + \xi^\alf \hat
d^\nu{}_{,\nu})_{,\alf} \equiv 0 $, which has to be considered
together with (\ref{LieLag}) (or (\ref{(+2+)})). This gives
additional contributions into the current (\ref{(+7+)}), also
(\ref{Cimath+}) or generalized (\ref{Cdagger}), and into the
superpotential (\ref{(+11+)}), also (\ref{Simath+}) or generalized
(\ref{Sdagger}):
 \bea
\hat \imath'^\alf &=&-\l[ \hat u'_\sig{}^\alf\xi^\sig + \hat
m'_\sig{}^{\alf\tau}\Bar D_\tau \xi^\sig \r]\,, \m{di}
\\
\hat \imath'^{\alf\beta}  &=&  - \hat
 m'_{\sig}{}^{[\alf\beta]}\xi^\sig \,
 \m {(+11+prime)}
 \eea
 where
 \be
\hat u'_\sig{}^\alf  = 2\BD_\beta (\delta^{[\alf}_\sig \hat
d^{\beta]} )\,,~~~~ \hat m'_\sig{}^{\alf\beta}=
 2 \delta^{[\alf}_\sig \hat d^{\beta]}\,,~~~~
\hat n'_\sig{}^{\alf\beta\gamma} =0\,.\m{umn}
 \ee
Note that a {\em construction} of these quantities does not depend
on the inner structure of $\hat d^{\nu}$.

\subsection{The Belinfante corrected family of covariant identically conserved quantities}
\m{sec:34}

Here, we modify the results of previous subsections by the use of
the Belinfante procedure \cite{Belinfante}. Using the general
Belinfante rule \cite{PK} we define a tensor density
 \be
\hat s^{\alf\beta\sig} \equiv - \hat s^{\beta\alf\sig} \equiv -
 \hat m_\lam{}^{\sig[\alf} \bar g^{\beta]\lam} -
 \hat m_\lam{}^{\alf[\sig} \bar g^{\beta]\lam} + \hat m_\lam{}^{\beta[\sig}
 \bar g^{\alf]\lam}\, ,
 \m{(+13+)}
 \ee
 which is called as a Belinfante correction,
add  $ \BD_{\beta}(\hat s^{\alf\beta\sig}\xi_{\sig})$ to both sides
of (\ref{(+10+)}) and obtain a new identity:
 \be
 \hat \imath^\alf_B
 \equiv \di_{\beta} \hat \imath^{\alf\beta}_B.
 \m{(+14+)}
 \ee
This modification cancels the spin term from the current
(\ref{(+7+A)}):
 \be
\hat \imath^\alf_B \equiv \l(- \hat u_\sig{}^{\alf} -\hat
n_\lam{}^{\alf\beta\gamma} \Bar R^\lam_{~\beta\gamma\sig} +
\BD_{\beta} \hat s^{\alf\beta}{}_{\sig}\r)  \xi^\sig+
 \hat z^{\alf}_B(\xi)  \equiv
\hat u_{B\sig}{}^\alf  \xi^\sig+
 \hat z^{\alf}_B(\xi),
 \m{(+15+)}
 \ee
a new $z$-term disappears also on Killing vectors of the background:
 \bea
 \hat z^{\alf}_B(\xi) &=&
\l(  \hat  m_\lam{}^{\beta\alf}  \bar g^{\tau\lam} +  \hat
m_\lam{}^{\alf\tau} \bar
 g^{\beta\lam} - \hat  m_\lam{}^{\tau\beta} \bar g^{\alf\lam}\r)
 \zeta_{\tau\beta}  \nonumber\\&+&
 \hat n_\lam{}^{\alf\tau\beta} \l(2\Bar D_{(\beta}
 \zeta_{\tau)}^\lam -
 \Bar D_{\sig} \zeta_{\beta\tau}\bar g^{\lam\sig}\r) .
 \m{Zmu-2}
 \eea
Thus, the  current $\hat i^\alf_B$ is defined, in fact, by the
modified energy-momentum tensor density
 $\hat u_{B\sig}{}^\alf$.
Because the new superpotential depends on  the $n$-coefficients
  only:
 \be
 \hat
 i^{\alf\beta}_B  \equiv 2
\l({\textstyle{1\over 3}}\BD_\rho \hat n_{\sig}{}^{[\alf\beta]\rho}+
\BD_\tau\hat
 n_{\lam}{}^{\tau\rho[\alf} \bar g^{\beta]\lam} \bar g_{\rho\sig}\r)
\xi^\sig - {\textstyle{4\over 3}} \hat
 n_\sig{}^{[\alf\beta]\lam}\BD_{\lam}\xi^\sig,
 \m{(+16+)}
 \ee
then due to the definition (\ref{(+5+)}) it vanishes for Lagrangians
with only the first order derivatives. On the other hand, the
superpotential (\ref{(+16+)}) is well adapted to theories with
second derivatives in Lagrangians, say, algebraically depending on
Riemanninan tensor.

It is important to  note that the Belinfante procedure cancels the
contributions of the divergence into currents and superpotentials,
see (\ref{di}) - (\ref{umn}). Let us show this. The Belinfante
correction constructed for the divergence by the rule (\ref{(+13+)})
for (\ref{umn}) is presented as
 \be
 \hat s'^{\alf\beta\sig}\xi_\sig =
 2\xi^{[\alf}\hat d^{\beta]}\, .
 \m{s-dm}
 \ee
Then, adding  $\BD_{\beta}(\hat s'^{\alf\beta\sig}\xi_{\sig})$ and
$\hat s'^{\alf\beta\sig}\xi_{\sig}$ to (\ref{di}) and
(\ref{(+11+prime)}) and keeping in mind (\ref{umn}) one obtains
easily
 \bea
\hat \imath'^\alf+\BD_{\beta}(\hat s'^{\alf\beta\sig}\xi_{\sig})
&=&-\l[ \hat u'_\sig{}^\alf\xi^\sig + \hat m'_\sig{}^{\alf\tau}\Bar
D_\tau \xi^\sig \r]+\BD_{\beta}(\hat s'^{\alf\beta\sig}\xi_{\sig}) =
0\,; \m{di=0}
\\
\hat \imath'^{\alf\beta}+\hat s'^{\alf\beta\sig}\xi_{\sig} &=&  -
\hat
 m'_{\sig}{}^{[\alf\beta]}\xi^\sig +\hat s'^{\alf\beta\sig}\xi_{\sig}=0\,
 \m {(+11+prime=0)}
 \eea

All the above can be applied exactly to the starred quantities in
Subsect.~\ref{sec:32}, obtaining in the result
 \be
 \hat \imath^{*\alf}_B
 \equiv \di_{\beta} \hat \imath^{*\alf\beta}_B
 \m{(+14+star)}
 \ee
where
 \bea
\hat \imath^{*\alf}_B &\equiv& \l(- \hat u^*_\sig{}^{\alf} -\hat
n^*_\lam{}^{\alf\beta\gamma} \Bar R^\lam_{~\beta\gamma\sig} +
\BD_{\beta} \hat s^{*\alf\beta}{}_{\sig}\r)  \xi^\sig+
 \hat z^{*\alf}_B(\xi),
 \m{(+15+star)}\\
\hat i^{*\alf\beta}_B  &\equiv& 2 \l({\textstyle{1\over 3}}\BD_\rho
\hat n^*_{\sig}{}^{[\alf\beta]\rho}+ \BD_\tau\hat
 n^*_{\lam}{}^{\tau\rho[\alf} \bar g^{\beta]\lam} \bar g_{\rho\sig}\r)
\xi^\sig - {\textstyle{4\over 3}} \hat
 n^*_\sig{}^{[\alf\beta]\lam}\BD_{\lam}\xi^\sig\,.
 \m{(+16+star)}
 \eea
The connection with the usual and starred Belinfante corrected
quantities is
 \bea \hat \imath^{*\alf}_B \equiv \hat
\imath^{\alf}_B &-& 2\BD_\beta\l[{{\di \lag} \over {\di
Q_{B;[\alf\beta]}}}\l.Q_{B}\r|^{\rho}_\sig \Bar D_\rho \xi^\sig
+\Bar D_{\rho} \l(\frac{\di\lag}{\di
Q_{B;[\sig\rho]}}\l.Q_{B}\r|^{[\alf}_\lam\bar
g^{\beta]\lam}\nonumber\r.\r.\\&+&\l.\l. \frac{\di\lag}{\di
Q_{B;[\alf\rho]}}\l.Q_{B}\r|^{[\sig}_\lam\bar
g^{\beta]\lam}-\frac{\di\lag}{\di
Q_{B;[\beta\rho]}}\l.Q_{B}\r|^{[\sig}_\lam\bar g^{\alf]\lam}
\r)\xi_\sig\r]\,,
\m{iBstar}\\
 \hat \imath^{*\alf\beta}_B \equiv \hat
\imath^{\alf\beta}_B &-& 2\l[{{\di \lag} \over {\di
Q_{B;[\alf\beta]}}}\l.Q_{B}\r|^{\rho}_\sig \Bar D_\rho \xi^\sig
+\Bar D_{\rho} \l(\frac{\di\lag}{\di
Q_{B;[\sig\rho]}}\l.Q_{B}\r|^{[\alf}_\lam\bar
g^{\beta]\lam}\nonumber\r.\r.\\&+&\l.\l. \frac{\di\lag}{\di
Q_{B;[\alf\rho]}}\l.Q_{B}\r|^{[\sig}_\lam\bar
g^{\beta]\lam}-\frac{\di\lag}{\di
Q_{B;[\beta\rho]}}\l.Q_{B}\r|^{[\sig}_\lam\bar g^{\alf]\lam}
\r)\xi_\sig\r].
 \m{iiBstar}
 \eea
The Belinfante corrected covariant conservation law and conserved
quantities for the united covariantized Lagrangian (\ref{p+q}) are
defined as
 \bea
 \hat\imath^{\dagger\alf}_B&\equiv& \di_\beta  \hat \imath^{\dagger\alf\beta}_B\,,
 \m{(+10+daggerB)} \\
\hat \imath^{\dagger\alf}_B &\equiv& p\, \hat\imath^{\alf}_B + q\,
\hat\imath^{*\alf}_B\,,
 \m{CdaggerB}\\
 \hat\imath^{\dagger\alf\beta}_B &\equiv& p\, \hat\imath^{\alf\beta}_B + q\,
 \hat\imath^{*\alf\beta}_B\,,
 \m{SdaggerB}
 \eea
presenting a Belinfante corrected family of identically conserved
quantities.

It was remarked earlier that the conserved quantities constructed in
\cite{Petrov2008,Petrov2009,Petrov2009cor,Petrov2011} are {\em
unique} for the Lagrangian and in the framework of the N{\oe}ther or
of the N{\oe}ther-Belinfante procedure. There is no a contradiction
with the results of this section where we suggest a {\em new} family
of conserved quantities. It is because we have found here various
possibilities to construct covariantized Lagrangians, in fact we
suggest a {\em family} of such Lagrangians (\ref{p+q}). Thus, for
each of the Lagrangians of the family the conserved quantities are
defined by an unique way.

\section{Conserved quantities for perturbations in arbitrary $D$-dimensional
metric theories}
 \m{sec:4}

\subsection{A metric theory} \m{sec:41}

 To present $D$-dimensional metric theory we consider the
Lagrangian:
 \be
 \hat L_D = - \frac{1}{2\k_D}\hat L_{g}(g_{\mu\nu}) + \hat
 L_{m}(g_{\mu\nu},\Phi)\,,
 \m{lag-g}
 \ee
which depends on the metric $g_{\mu\nu}$ and $\Phi$ and their
derivatives up to a second order, where  $\Phi$ defines matter
sources without concretization. Thus $\hat L_{g}$ can be thought as
an algebraic function of the metric and Riemannian tensors, $\hat
L_{g}(g_{\mu\nu})= \hat L_{g}(g_{\mu\nu},
R^\alf{}_{\rho\beta\sig})$, that can be arbitrary. Variation of
(\ref{lag-g}) with respect to $g^{\mu\nu}$ leads to the
gravitational equations:
 \be
\hat{\cal G}_{\mu\nu} = \k_D\hat T_{\mu\nu}\, . \m{ddd+}
 \ee
Variation of (\ref{lag-g}) with respect to $\Phi $ gives
corresponding matter equations. Below we will use also the
background Lagrangian defined as $\Bar {\hat L}_D = \hat L_D(\bar
g_{\mu\nu},\Bar\Phi)$ and corresponding background gravitational
equations
 \be
\Bar{\hat{\cal G}}_{\mu\nu} = \k_D\Bar{\hat T}_{\mu\nu}\,
\m{ddd+back}
 \ee
and matter equations. We set that the background fields $\bar
g_{\mu\nu}$ and $\Bar\Phi$ satisfy the background equations and,
thus, are known (fixed).

In the present subsection, the subject of our attention is the
gravitational part of the Lagrangian (\ref{lag-g}). Basing on the
results of previous section, we set $Q^A = \{g_{\mu\nu}\}$ and
incorporate an external metric $\Bar g_{\mu\nu}$ into $\hat L_g$ in
(\ref{lag-g}). A presentation of the Lagrangian in an ``explicitly''
covariant form with the use of $\Bar g_{\mu\nu}$ is carried out
exactly by the recipe of previous section. We change partial
derivatives by covariant derivatives defined with respect to $\Bar
g_{\mu\nu}$. Thus, we transform the pure metric Lagrangian $\hat
L_g$ into an explicitly covariant form:
 \be
\hat L_g = \lag_g = {\lag}_g (g_{\mu\nu},\, g_{\mu\nu;\alf},\,
g_{\mu\nu;\alf\beta},\,\Bar g_{\mu\nu},\,\Bar
R^\lam{}_{\tau\rho\sig}).
 \m{Lg=Lc}
 \ee

Now we derive the coefficients (\ref{(+5+)+}), (\ref{m(+4+)}) and
(\ref{u(+3+)}) for the Lagrangian  $\lag = -\lag_g/2\k_D $, setting
there $Q^A = \{g_{\mu\nu}\}$. Thus, the coefficients are defined as
follows. We directly rewrite $n$ and $m$ coefficients:
  \be \hat n_\sig{}^{\alf\tau\beta} = -\frac{1}{4\k_D}
\l[{{\di \lag_g} \over {\di  g_{\mu\nu;\beta\alf}}}
 \l.g_{\mu\nu}\r|^\tau_\sig +
 {{\di \lag_g} \over {\di g_{\mu\nu;\tau\alf}}}
 \l.g_{\mu\nu}\r|^\beta_\sig\r].
\m{(+5+g)}
 \ee
 \be
  \hat m_\sig{}^{\alf\tau} = -\frac{1}{2\k_D}\l[
{{\delta \lag_g} \over {\delta g_{\mu\nu;\alf}}}
 \l.g_{\mu\nu}\r|^\tau_\sig -
 {{\di \lag_g} \over {\di g_{\mu\nu;\tau\alf}}}
\Bar D_\sig g_{\mu\nu} +
 {{\di \lag_g} \over {\di  g_{\mu\nu;\beta\alf}}}
 \Bar D_\beta (\l.g_{\mu\nu}\r|^\tau_\sig)\r]\, ,
\m{(+4+g)}
 \ee
where $\l. g_{\mu\nu} \r|^\alf_\beta=
-2g_{\beta(\mu}\delta^\alf_{\nu)}$ and
 \be
 {{\delta \lag_g} \over {\delta  g_{\mu\nu;\alf}}}\equiv {{\di \lag_g} \over {\di
g_{\mu\nu;\alf}}} -
 \Bar D_\beta
\l({{\di \lag_g} \over {\di  g_{\mu\nu;\alf\beta}}}\r)\,.
 \m{Lagrangian}
 \ee
It is useful to present $u$ coefficient in a structured form:
  \be
 \hat u_\sig{}^\alf =  -\frac{1}{\k_D}\l[\hat {\cal G}_\sig^\alf +
 \k_D \hat{\cal U}_{\sig}{}^\alf{}+\k_D \hat n_\lam{}^{\alf\tau\beta}\Bar
R^\lam_{~\tau\beta\sig}\r]\, , \m{(+3+g)}
 \ee
we use the notations
 \bea
\hat {\cal G}^\alf_\sig &\equiv & \frac{1}{2}{{\delta \lag_g} \over
{\delta g_{\mu\nu}}} \l.g_{\mu\nu}\r|^\alf_\sig \equiv - {{\delta
\lag_g} \over {\delta g_{\mu\alf}}} g_{\mu\sig} \equiv {{\delta
\lag_g} \over {\delta
g^{\mu\sig}}} g^{\mu\alf}\,,\m{calGmunu}\\
\hat{\cal U}_{\sig}{}^\alf{}
  & \equiv & -\frac{1}{2\k_D}\l( {{\di \lag_{g} } \over
{\di g_{\mu\nu;\beta\alf}}}\Bar D_{\sig\beta} g_{\mu\nu} +{{\delta
\lag_{g} } \over {\delta g_{\mu\nu;\alf}}}\Bar D_\sig g_{\mu\nu} -
\delta^\alf_\sig \lag_{g} \r) \,. \m{TG}
 \eea
 As usual, ${{\delta \lag_g}/{\delta g_{\mu\nu}}}$ means Lagrangian
derivatives, see (\ref{L-derivativeCov}), $\hat {\cal
G}^\alf{}_\sig$ is exactly the symmetrical left hand side of
(\ref{ddd+}), and $\hat {\cal U}^\alf{}_\sig$ is the generalized
canonical energy-momentum related to the gravitational Lagrangian
(\ref{Lg=Lc}).

\subsection{Currents and superpotentials for perturbations}
 \m{sec:42}

Incorporation of the background metric is a key point, basing on
which one has  a possibility to describe perturbations.
Perturbations are determined by the way when a one solution
(dynamical) of the theory is considered as a perturbed system with
respect to another solution (background) of the same theory. Then
the background spacetime acquires a {\em real} sense, not auxiliary.
Perturbations in such a derivation are exact (not infinitesimal or
approximate), and then linear or of higher order approximations
follow easily. We will denote $\delta$ as an exact difference
between dynamical and background quantities:
 \be \delta Q^A = Q^A- \Bar{ Q^A}.
 \m{deltaQ}
 \ee
Following to the Katz-Bi\v c\'ak-Lynden-Bell ideology \cite{KBL} we
construct the metric Lagrangian for perturbations:
 \be
 \lag_{G} = -\frac{1}{2\k_D}\l(\lag_{g} - \Bar\lag_{g} + \di_\alf \hat d^\alf\r)\, .
 \m{ArbitraryLagPert}
 \ee
By the definition, the Lagrangian has to vanish for vanishing
perturbations, therefore usually $\hat d^\alf$ disappears for
vanishing perturbations.

At first we construct the N{\oe}ther canonical conserved quantities.
Substituting (\ref{(+5+g)}), (\ref{(+4+g)}) and (\ref{(+3+g)}) into
(\ref{(+7+)}) (or to (\ref{(+7+A)})), applying the barred procedure,
subtracting one from another  and taking into account the divergence
in the way (\ref{di}) - (\ref{umn}), one obtains the current
corresponding to (\ref{ArbitraryLagPert}):
 $
\delta\hat \imath^\alf = \hat \imath^\alf - \Bar{\hat \imath^\alf}
+\hat \imath'^\alf
 $.
Then we use the dynamical equations (\ref{ddd+}) in $\hat
u_{\sig}{}^{\alf}$. We change $\hat{\cal G}_{\mu\nu}$ (as a part of
$\hat u_{\sig}{}^{\alf}$, see (\ref{(+3+g)})) by the matter
energy-momentum ${\hat T_{\mu\nu}}$ at the right hand side of
(\ref{ddd+}). Next, we do the same combining $\Bar{\hat
u}_{\sig}{}^{\alf}$ and the barred equations (\ref{ddd+back}). In
the result one obtains that the identically conserved current
$\delta\hat \imath^\alf$ related to (\ref{ArbitraryLagPert})
transforms into the current for perturbations:
 \be
\hat {\cal I}^\alf(\xi) = {}\hat {\Theta}_{\sig}{}^\alf \xi^\sig +
{}\hat {\cal M}^{\sig\alf\beta}\di_{[\sig}\xi_{\beta]} + {}\hat
{\cal Z}^\alf(\xi)\,.
  \m{(+7+A+)}
 \ee
Now, the conservation law:
 \be
 \di_\alf \hat {\cal I}^\alf(\xi) = 0\,
 \m{identity-Ic}
 \ee
takes a place due to the field equations, {\em not identically}. The
generalized canonical energy-momentum tensor density, spin-term and
$Z$-term for perturbations are
 \bea
{}\hat {\Theta}_{\sig}{}^\alf &\equiv &  \delta\hat T^\alf_\sig +
\delta \hat{\cal U}_{\sig}{}^\alf +
\k^{-1}_D\BD_\beta(\delta^{[\alf}_\sig\hat d^{\beta]}) \,,
\m{7A'}\\
\hat {\cal M}^{\sig\alf\beta} &\equiv & \delta\hat
 m_{\rho}{}^{\alf\beta}\bar g^{\sig\rho}- \k^{-1}_D\bar g^{\sig[\alf}\hat
 d^{\beta]}\,,
 \m{7A''}\\
\hat {\cal Z}^\alf(\xi) &\equiv &
  -\delta \hat z^{\alf} + \k^{-1}_D\zeta^{[\alf}_\beta\hat d^{\beta]}
\,
 \m{7A'''}
 \eea
 where $\delta$ has a sense of a general definition
 (\ref{deltaQ}).
To present the value of the current (\ref{(+7+A+)}), one has to take
solutions to the equations  (\ref{ddd+}) and (\ref{ddd+back}) and
use them for calculating concrete values of the quantities
(\ref{(+5+g)}), (\ref{(+4+g)}), (\ref{TG}) and (\ref{(+8+)}).
Starting from (\ref{(+11+)}), by the same way we construct a
superpotential corresponding to the current (\ref{(+7+A+)}):
\hspace*{-1.0cm}  \bea  \delta\hat \imath^{\alf\beta} &=& \hat
\imath^{\alf\beta} - \Bar{\hat \imath^{\alf\beta}} +\hat
\imath'^{\alf\beta}
 \goto\nonumber\\
 \hat {\cal I}^{\alf\beta}  &=& \l({\textstyle{2\over 3}}
 \BD_\lam  \delta \hat n_{\sig}{}^{[\alf\beta]\lam}  - \delta\hat
 m_{\sig}{}^{[\alf\beta]} - 2 \k^{-1}_D\delta^{[\alf}_\sig \hat d^{\beta]}\r)\xi^\sig   -
 {\textstyle{4\over 3}} \delta\hat n_{\sig}{}^{[\alf\beta]\lam}
 \BD_\lam  \xi^\sig.
 \m {(+11+A)}
 \eea
Then, instead of  (\ref{identity-Ic}) one can use the conservation
law in the form:
 \be
\hat {\cal I}^{\alf}(\xi)= \di_\alf \hat {\cal I}^{\alf\beta}(\xi)
\,.
 \m{calCLcanonical}
 \ee
It is not identity, but the conservation law for perturbations
determined by the solutions to the equations (\ref{ddd+}) and
(\ref{ddd+back}).

Analogously the starred conservation law can be constructed. Thus,
more generally, the family of the N{\oe}ther canonical conservation
laws for perturbations corresponding to the presentation (\ref{p+q})
- (\ref{Sdagger}) has a form:
 \be
\hat {\cal I}^{\dagger\alf}(\xi)= \di_\alf \hat {\cal
I}^{\dagger\alf\beta}(\xi) \,.
 \m{calCLcanonicalD}
 \ee
It is edifying to find a connection of the generalized conserved
quantities in (\ref{calCLcanonicalD}) with known ones in 4D GR.
Note, that even in D-dimensional GR there is no a difference between
starred and non-starred quantities. It is because the Einstein part
of (\ref{DiLDDgmn}) in Appendix \ref{Appendix2} does not contain an
antisymmetrical part. Therefore for GR the conservation law
(\ref{calCLcanonicalD}) is a single one, not a family. Thus, it is
enough to take into account the Einstein part in (\ref{CanSupEGB}) -
(\ref{divd})  in Appendix \ref{Appendix2} to derive a superpotential
in (\ref{calCLcanonicalD}) for GR. Next, choosing the Minkowski
background in Cartesian coordinates and the translation Killing
vectors in the form $\xi^\alf= \delta^\alf_{(\beta)}$ one recognizes
that the superpotential in (\ref{calCLcanonicalD}) goes to the very
known Freud superpotential \cite{Freud39} in 4D GR.

To construct the Belinfante corrected conserved currents for the
perturbed system (\ref{ArbitraryLagPert}) we substitute
(\ref{(+5+g)}), (\ref{(+4+g)}) and (\ref{(+3+g)}) into
(\ref{(+15+)}) and subtract the corresponding barred expression
(\ref{(+15+)}):
  $
 \delta\hat \imath^\alf_{B}= \hat \imath^\alf_{B} - \Bar{\hat \imath^\alf_{B}}
  $.
The same is obtained after applying the N{\oe}ther-Belinfante method
directly to the Lagrangian in (\ref{ArbitraryLagPert}). Again, using
the equations (\ref{ddd+}) and their barred version (\ref{ddd+back})
in $\hat u_{\sig}{}^{\alf}$ and $\Bar{\hat u}_{\sig}{}^{\alf}$, the
current $ \delta\hat \imath^\alf_{B}$ related to
(\ref{ArbitraryLagPert}) transforms into
 \be
\hat {\cal I}_B^\alf(\xi) = \hat{\Theta}_{B\sig}{}^\alf\xi^\sig +
\hat {\cal Z}_B^\alf(\xi) \, .
 \m{(+15+A)}
 \ee
Thus, one has a conservation law for perturbations
 \be
 \di_\alf \hat {\cal I}_B^\alf(\xi) = 0\,,
 \m{identity-Ib}
 \ee
which takes a place on the field equations, {\em not identically}.
Of course, the current (\ref{(+15+A)}) does not contain a spin term,
unlike (\ref{(+7+A+)}). The Belinfante corrected energy-momentum
tensor density and $Z$-term for perturbations are
 \bea
\hat{\Theta}_{B\sig}{}^\alf &\equiv & \delta\hat T^\alf_\sig +
\delta \hat{\cal U}_{\sig}{}^\alf  + \BD_{\beta}\delta \hat
s^{\alf\beta}{}_{\sig} \, ,
\m{7A+}\\
\hat {\cal Z}_B^\alf(\xi) &\equiv & -\delta\hat z_B^{\alf}(\xi)\, .
 \m{(7A++)}
 \eea
Starting from (\ref{(+16+)}), a superpotential corresponding to the
current (\ref{(+15+A)}) is constructed analogously:
  \bea
  \delta\hat \imath^{\alf\beta}_{B}&=& \hat \imath^{\alf\beta}_{B} - \Bar{\hat \imath^{\alf\beta}_{B}} \goto\nonumber\\
\hat {\cal I}_B^{\alf\beta} & =& \equiv 2 \l({\textstyle{1\over
3}}\BD_\rho \delta\hat n_{\sig}{}^{[\alf\beta]\rho}+
\BD_\tau\delta\hat
 n_{\lam}{}^{\tau\rho[\alf} \bar g^{\beta]\lam} \bar g_{\rho\sig}\r)
\xi^\sig - {\textstyle{4\over 3}} \delta\hat
 n_\sig{}^{[\alf\beta]\lam}\BD_{\lam}\xi^\sig.
 \m{(+16+A)}
 \eea
Instead of (\ref{identity-Ib}) the conservation law for the
perturbations can be presented also in the form:
 \be
\hat {\cal I}_B^{\alf}(\xi)=  \di_\alf \hat {\cal
I}_B^{\alf\beta}(\xi) \,.
 \m{calCLBelinfante}
 \ee
It is not identity, all the quantities (\ref{(+5+g)}),
(\ref{(+4+g)}), (\ref{TG}), also (\ref{(+13+)}) and (\ref{Zmu-2})
are determined by the solutions to the equations (\ref{ddd+}) and
(\ref{ddd+back}).

By the same way the starred Belinfante corrected conservation law
can be constructed. More generally, the family of the conservation
laws corresponding to the presentation  (\ref{p+q}),
(\ref{(+10+daggerB)}) - (\ref{SdaggerB}) takes a place:
 \be
\hat {\cal I}_B^{\dagger\alf}(\xi)=  \di_\alf \hat {\cal
I}_B^{\dagger\alf\beta}(\xi) \,.
 \m{calCLBelinfanteD}
 \ee
Again, for GR the conservation law (\ref{calCLBelinfanteD}) is a
single one, not a family. Using the Einstein part in
(\ref{BelSupEGB}) and (\ref{BarBelSupEGB})  in Appendix
\ref{Appendix2} one can be convinced that on the Minkowski
background in Cartesian coordinates and with the translation Killing
vectors the superpotential in (\ref{calCLBelinfanteD}) transforms
into the well known Papapetrou superpotential \cite{Papapetrou48} in
4D GR, see also \cite{PK}.

\section{Applications}
\m{sec:5} 

\subsection{Conserved charges in EGB gravity}
\m{sec:51}

In this section, we apply the results of previous sections to
calculate mass of the Schwarz\-schild-anti-de Sitter (S-AdS) BH
\cite{BD+} in the EGB gravity. Reasons why we have chosen this
solution are as follows. First, in the Einstein theory there are no
differences between various conserved quantities of the new family
(see discussion in previous section), but they are exist in the EGB
gravity, which is a one of the most popular modifications of GR.
Second, the S-AdS BH in the EGB gravity is a more known solution,
which is frequently used as a standard solution in applications.

Historically, in
\cite{Petrov2009,Petrov2009cor,Petrov2009a,Petrov2011}, it was
suggested the {\em starred} variant of conservation laws for
perturbations based on the identities (\ref{(+10+star)}) and
(\ref{(+14+star)}) only. In  in Appendix \ref{Appendix2}, we present
the necessary formulae for EGB gravity including the starred
superpotentials (\ref{CanSupEGB}) - (\ref{BarBelSupEGB}) related to
(\ref{(+11+star)}) and (\ref{(+16+star)}) on the right hand sides of
(\ref{(+10+star)}) and (\ref{(+14+star)}). In
\cite{Petrov2009,Petrov2009cor}, basing on these superpotentials,
already we have obtained the mass for the aforementioned solution,
which coincides exactly with the standard results. In previous
sections, expanding the results of
\cite{Petrov2009,Petrov2009cor,Petrov2009a,Petrov2011}, we have
suggested not only starred conserved quantities, but the family of
conserved quantities united by (\ref{calCLcanonicalD}) and
(\ref{calCLBelinfanteD}). Here, we test all of them, calculating
mass for the S-AdS BH in EGB gravity with the use of the generalized
superpotentials in (\ref{calCLcanonicalD}) and
(\ref{calCLBelinfanteD}).

Rewrite the conservation laws of all the types for perturbations in
the united form:
  \be
 \hat {\cal I}^\alf_D(\xi) = \di_\beta \hat {\cal I}^{\alf\beta}_D (\xi)\,.
 \m{generalCLs}
  \ee
Here, a superpotential can be one of the set $ \hat {\cal
I}^{\alf\beta}_D= \{\hat {\cal I}^{\dagger\alf\beta},\,\hat {\cal
I}_B^{\dagger\alf\beta} \}$; the same is related to the currents for
perturbations. The conservation law (\ref{generalCLs}) allows us to
construct the conserved charges in generalized form in
$D$-dimensions:
 \be
{\cal P}(\xi) = \int_\Sigma d^{D-1} x\,\hat {\cal I}^0_D(\xi) =
\oint_{\di\Sigma} dS_i \,\hat {\cal I}^{0i}_D(\xi)\,.
 \m{charges}
 \ee
 In next subsection we concretize the notations
for a section $\Sigma$ and its boundary $\di\Sigma$.

With the use of (\ref{Sdagger}) and (\ref{SdaggerB}),
(\ref{(+11+A)}) and (\ref{(+16+A)}), and starred (\ref{(+11+A)}) and
(\ref{(+16+A)}) we represent the superpotentials in
(\ref{calCLcanonicalD}) and (\ref{calCLBelinfanteD}) in the form:
 \bea
\hat{\cal I}^{\dagger\alf\beta} &=&\hat{\cal I}^{*\alf\beta} +
p\l[(\hat\imath^{\alf\beta}- \hat\imath^{*\alf\beta})-
(\Bar{\hat\imath^{\alf\beta}}-\Bar{\hat\imath^{*\alf\beta}})\r]
=\hat{\cal I}^{*\alf\beta}+p\Delta
\hat{\cal I}^{\alf\beta}\,, \m{canonicalDIF}\\
\hat{\cal I}^{\dagger\alf\beta}_B &=&\hat{\cal I}_B^{*\alf\beta} +
p\l[\l(\hat\imath_B^{\alf\beta}- \hat\imath_B^{*\alf\beta}\r)-
\l(\Bar{\hat\imath_B^{\alf\beta}}-\Bar{\hat\imath_B^{*\alf\beta}}\r)\r]=
\hat{\cal I}_B^{*\alf\beta} +p\Delta \hat{\cal I}_B^{\alf\beta}\,.
\m{BelinfanteDIF}
 \eea
The expressions in square brackets are defined by the formulae
(\ref{Simath+}) and (\ref{iiBstar}). Because for the S-AdS BH  the
starred superpotentials already have been checked we need only to
calculate
  \be
\Delta{\cal P}(\xi) = p\oint_{\di\Sigma} dS_i \,\Delta\hat {\cal
I}^{0i}_D(\xi)\,
 \m{chargesDelta}
 \ee
where $ \Delta\hat {\cal I}^{\alf\beta}_D= \{ \Delta\hat {\cal
I}^{\alf\beta},\, \Delta\hat {\cal I}_B^{\alf\beta}\}$.

To define $ \Delta\hat {\cal I}^{\alf\beta}_D$ in
(\ref{canonicalDIF}) and (\ref{BelinfanteDIF}) for the EGB gravity
one needs in an antisymmetrical in $\alf$ and $\beta$ part of the
expression (\ref{DiLDDgmn}), see (\ref{Simath+}) and
(\ref{iiBstar}). It is
 \be
\frac{\di \lag_{EGB}}{\di
g_{\mu\nu;[\alf\beta]}}=\frac{2\alf\sqrt{-g}}{\k_D}\l(
R^{\alf(\mu}g^{\nu)\beta}- R^{\beta(\mu}g^{\nu)\alf}\r).
 \m{DiantiAB}
 \ee
 Thus, from (\ref{Simath+}) one has
 \bea
\hat\imath^{\alf\beta}- \hat\imath^{*\alf\beta} &=& 2\frac{\di
\lag_{EGB}}{\di g_{\mu\nu;[\alf\beta]}}\pounds_\xi
g_{\mu\nu}\nonumber\\&=&-
\frac{8\alf\sqrt{-g}}{\k_D}\l(\BD_\mu\xi^\rho
+\xi^\tau\Delta^\rho_{\tau\mu}\r)\l(R^{[\alf}_\rho g^{\beta]\mu}+
R^{\mu[\alf}\delta^{\beta]}_{\rho}\r)\nonumber\\ &=& -
\frac{8\alf\sqrt{-g}}{\k_D}\l(R^{[\alf}_\rho g^{\beta]\mu}+
R^{\mu[\alf}\delta^{\beta]}_{\rho}\r)D_\mu\xi^\rho \, ,
 \m{i-i}
 \eea
where the quantities $\Delta^\rho_{\tau\mu}$ are presented by the
relation (\ref{DeltaDefD})  in Appendix \ref{Appendix2}, and the
definitions (\ref{LieQ+Ad}) and (\ref{LieQ+Add}) are used also. From
(\ref{iiBstar}) one has
 \bea
\hat\imath_B^{\alf\beta}- \hat\imath_B^{*\alf\beta} &=& 2\l[{{\di
\lag} \over {\di g_{\mu\nu;[\alf\beta]}}}
\l.g_{\mu\nu}\r|^{\rho}_\sig \Bar D_\rho \xi^\sig  +\Bar D_{\rho}
\l(\frac{\di\lag}{\di
g_{\mu\nu;[\sig\rho]}}\l.g_{\mu\nu}\r|^{[\alf}_\lam\bar
g^{\beta]\lam}\r.\r. \nonumber\\&+& \l.\l. \frac{\di\lag}{\di
g_{\mu\nu;[\alf\rho]}}\l.g_{\mu\nu}\r|^{[\sig}_\lam\bar
g^{\beta]\lam}-\frac{\di\lag}{\di
g_{\mu\nu;[\beta\rho]}}\l.g_{\mu\nu}\r|^{[\sig}_\lam\bar
g^{\alf]\lam}   \r)\xi_\sig\r] \nonumber\\&=&
\frac{4\alf\sqrt{-g}}{\k_D}\l\{-2\l(R^{[\alf}_\sig g^{\beta]\rho}+
R^{\rho[\alf}\delta^{\beta]}_{\sig}\r)\BD_\rho\xi^\sig\r.
\nonumber\\&+& \l.\xi_\sig\BD_\rho\l[R^\sig_\lam g^{\rho[\alf}\Bar
g^{\beta]\lam} + g^{\sig\rho}R^{[\alf}_\lam\Bar g^{\beta]\lam} -
R^{[\alf}_\lam g^{\beta]\rho}\Bar g^{\sig\lam}\r.\r.
\nonumber\\&+&\l.\l. 2\l( R^{\sig[\alf}\Bar g^{\beta]\rho} -
R^{\rho[\alf}\Bar g^{\beta]\sig} -R^\rho_\lam g^{\sig[\alf}\Bar
g^{\beta]\lam} \r) \r]\r\}.\m{iB-iB}
 \eea
 The barred expressions are obtained easily.

\subsection{Mass of the Schwarzschild-AdS black hole in EGB gravity}
\m{sec:52}

The S-AdS solution in EGB gravity (see formulae (\ref{EGBaction}),
(\ref{GB-term}) in Appendix \ref{Appendix2}) has a form \cite{BD+}:
  \be
 d s^2 = -fdt^2 + f^{-1}dr^2 +
 r^2\sum_{a,b}^{D-2}q_{ab}dx^adx^b\,
 \m{S-AdS}
 \ee
 with the metric components $g_{00}=-f(r)$ and $g_{11}=f^{-1}(r)$ where
 \bea \hspace*{-0.5cm}
 f(r)& =&  1+\frac{r^2}{2\alf(D-3)(D-4)}\nonumber\\&\times&
\l\{1 \pm \sqrt{1 - \frac{4\Lambda_{0}}{\Lambda_{EGB}} +
4\alf(D-3)(D-4)\frac{r_0^{D-3}}{r^{D-1}}} \r\},
 \label{39bis}\\
  \Lambda_{EGB}& =& -\frac
{(D-2)(D-1)}{{2\alf}(D-4)(D-3)}\,.
 \m{LambdaEGB}
 \eea
The last term in (\ref{S-AdS}) describes $(D-2)$-dimensional sphere
of  the radius $r$, and $q_{ab}$ depends on coordinates on the
sphere only. The Christoffel symbols corresponding (\ref{S-AdS}) are
 \be
 \Gamma^1_{00} = \frac{ f { f}'}{2} ,\qquad
 \Gamma^0_{10} = \frac{{ f}'}{2{f}} ,\qquad
 \Gamma^1_{11} = -\frac{{f}'}{2{f}} ,\qquad
 \Gamma^a_{1b} = \frac{1}{r}\, \delta^a_b ,\qquad
 \Gamma^1_{ab} = -rf q_{ab}\,
 \m{ChristAdS}
 \ee
where `prime' means $\di/\di r$. The Riemannian, Ricci tensors and
curvature scalar corresponding to (\ref{S-AdS}) are
 \bea
R_{0101}&=& \half f''\,, \qquad R_{0a0b}= \half rff'q_{ab}\,
,\qquad  R_{1a1b}= - \frac{rf'}{2f}q_{ab}\, ,\nonumber\\
R_{abcd}&=& - r^2(f-1) (q_{ac}q_{bd}-q_{ad}q_{bc})\, ;
 \nonumber\\
R_{00} &=& \frac{f}{2}\l(f'' + f'\frac{D-2}{r}\r)\, ,\qquad R_{11}
= - \frac{1}{2f}\l(f'' + f'\frac{D-2}{r}\r)\, ,\nonumber\\
R_{ab} &=& -\l[(f-1)(D-3)+ rf'\r]q_{ab}\, ; \nonumber\\
R &=& -\l(f'' + f'\frac{D-2}{r}\r)-\frac{D-2}{r^2}\l[(f-1)(D-3)+
rf'\r]\, .
  \m{SAdS-R}
 \eea

The barred solution (\ref{S-AdS}) is defined by the condition
$r_0=0$ and presents a background AdS solution with
 \be
\Bar f(r) = 1- r^2\frac{2\Lambda_{eff}}{(D-1)(D-2)}\, .
 \m{fBar}
 \ee
The effective cosmological constant is defined as
 \be
 \Lambda_{eff} = \frac{\Lambda_{EGB}}{2} \l(1\pm \sqrt{1 -
 \frac{4\Lambda_{0}}{\Lambda_{EGB}}}\r)\,,
 \label{Leff+}
 \ee
thus $\Lambda_{eff}$ is negative, see (\ref{LambdaEGB}). For the
metric (\ref{S-AdS}) the relation  $\sqrt{-g_D} = \sqrt{-\Bar g_D} =
r^{D-2}\sqrt{\det q_{ab}}$ has a place and is important for
calculations. The background Christoffel symbols are the barred
expressions (\ref{ChristAdS}). The background Riemannian, Ricci
tensors and curvature scalar are
  \be \Bar R_{\mu\alf\nu\beta} = 2\Lambda_{eff}\frac{(\Bar
g_{\mu\nu}\Bar g_{\alf\beta} - \Bar g_{\mu\beta}\Bar
g_{\nu\alf})}{(D-2)(D-1)},~ \Bar R_{\mu\nu} =
2\Lambda_{eff}\frac{\Bar g_{\mu\nu}}{D-2},~ \Bar R =
2\Lambda_{eff}\frac{D}{D-2}\, . \label{R}
 \ee
The barred expressions (\ref{SAdS-R}) go to (\ref{R}). It is
evidently that for solutions
 (\ref{S-AdS}) perturbations can be described only by $\Delta f = f-\Bar f$.
Keeping only the first order term with respect to $1/r$, we obtain
 \be
 \Delta f = \pm \l(\sqrt{1 - \frac{4\Lambda_{0}}{\Lambda_{EGB}}}\r)^{-1}
\l(\frac{r_0}{r}\r)^{D-3}\, .
 \m{Deltaf}
 \ee

In \cite{Petrov2009,Petrov2009cor}, the mass of the S-AdS BH has
been obtained with the use of the starred conserved quantities. The
starred superpotentials (\ref{(+11+A)}) or (\ref{(+16+A)}) in the
EGB gravity are presented by (\ref{CanSupEGB}) -
(\ref{BarBelSupEGB}). Thus, using a timelike Killing vector $\xi^\mu
= (-1,\, {\bf 0})$ of the background and the S-AdS BH data
(\ref{S-AdS}) - (\ref{Deltaf}), one obtains
   \be
M = \lim_{r\goto\infty}\oint_{\di\Sigma} dx^{D-2} \,\sqrt{-\Bar g_D}
{\cal I}^{01}_D(\xi)= \frac{(D-2)r_0^{D-3}}{4G_D}\,.
 \m{chargesSAdS}
 \ee
Here,  $ \hat {\cal I}^{\alf\beta}_D=\{\hat {\cal
I}^{*\alf\beta},\,\hat {\cal I}_B^{*\alf\beta}\}$; $\di\Sigma$ is
the  $(D-2)$ dimensional boundary of $\Sigma$ that is a spacelike
$(D-1)$ hypersurface $x^0 = \const$. The result (\ref{chargesSAdS})
is the standard accepted result obtained with using the various
approaches (see, e.g., \cite{DT2,DerKatzOgushi},
\cite{Paddila,AFrancavigliaR,Okuyama} and references therein).

To examine the {\em family} of conservation laws
(\ref{calCLcanonicalD}) and (\ref{calCLBelinfanteD}), calculating
the mass of the S-AdS BH and keeping in mind (\ref{chargesSAdS}), it
is enough to calculate the integral (\ref{chargesDelta})
   \be
\Delta M = p\lim_{r\goto\infty}\oint_{\di\Sigma} dx^{D-2}
\,\sqrt{-\Bar g_D} \Delta{\cal I}^{01}_D(\xi)\,.
 \m{chargesSAdSDelta}
 \ee
Here, the Killing vector $\xi^\mu = (-1,\, {\bf 0})$ is used again,
and (\ref{i-i}) and (\ref{iB-iB}) have to be calculated for
(\ref{S-AdS}) - (\ref{Deltaf}). In the canonical N{\oe}ther case one
obtains $\hat\imath^{\alf\beta}- \hat\imath^{*\alf\beta}\equiv 0$.
The calculations for the Belinfante corrected case give
 \be
\hat\imath^{\alf\beta}_B- \hat\imath^{*\alf\beta}_B =
\frac{4\alf\sqrt{-\Bar g}}{\k_D}\l(g^{00}- \Bar g^{00}\r) \l(
g^{ab}R^0_0-R^{ab}\r)\Bar \Gamma^1_{ab}\,.
 \m{iB-iBstar}
 \ee
The formulae (\ref{S-AdS}) - (\ref{Deltaf}) show that
$\Delta\hat{\cal I}^{01}_B =\hat\imath^{\alf\beta}_B-
\hat\imath^{*\alf\beta}_B\sim 1/r^{D+1}$. Thus, for the canonical
N{\oe}ther case (\ref{calCLcanonicalD}) $\Delta M=0$ identically,
whereas for the Belinfante corrected case (\ref{calCLBelinfanteD})
$\Delta M=0$ due to the asymptotical behavior. This means that,
calculating mass of the S-AdS BH in the EGB gravity, all the
superpotentials of the new family give the same standard result
(\ref{chargesSAdS}).

\section{Concluding remarks}
\m{sec:6} 

In the paper, expanding possibilities for constructing conservation
laws and conserved quantities for perturbations on arbitrary curved
backgrounds in metric theories, we have suggested a new family of
such expressions and quantities. Particular types of conservation
laws, which relate to this family, already exist and have been
applied. Thus, in
\cite{Petrov2009,Petrov2009cor,Petrov2009a,Petrov2011}, the
conserved quantities denoted here as a starred ones have been
presented, see Subsect.~\ref{sec:32} and Appendix \ref{Appendix2}.
The quantities derived in \cite{DerKatzOgushi} are related to the
type considered in Subsect.~\ref{sec:31}. Let us show this.

Reformulating the superpotential (4.8) in the Deruelle, Katz and
Ogushi paper \cite{DerKatzOgushi} constructed in the EGB gravity in
our notations, one obtains
 \be
 {\hat \imath^{\alf\beta}_{DKO}}  =
{\sqrt{-g}\over \k_D}  {D^{[\alf} \xi^{\beta]}}
 -
\frac{2\alf\sqrt{-g}}{\k_D}\l\{{ R}_\sig{}^{\lam\alf\beta} +
 4 {R^{\lam[\alf} \delta^{\beta]}_\sig} + \delta_\sig^{[\alf}
 {g^{\beta]\lam} R}
 \r\}D_\lam \xi^\sig\, .
 \m{CanSupEGBDKO}
 \ee
Reformulating the starred superpotential (\ref{CanSupEGB}) in the
terms of the dynamical covariant derivative (\ref{LieQ+Ad}) with the
use of (\ref{LieQ+Add}), one obtains
 \be
 {\hat \imath^{*\alf\beta}}  =
{\sqrt{-g}\over \k_D}  {D^{[\alf} \xi^{\beta]}}
 -
\frac{2\alf\sqrt{-g}}{\k_D}\l\{{ R}_\sig{}^{\lam\alf\beta} +
 4 {g^{\lam[\alf} R^{\beta]}_\sig} + \delta_\sig^{[\alf}
 {g^{\beta]\lam} R}
 \r\}D_\lam \xi^\sig\, .
 \m{CanSupEGB-D}
 \ee
Comparing ${\hat \imath^{\alf\beta}_{DKO}} -  {\hat
\imath^{*\alf\beta}} $ and the difference ${\hat
\imath^{\alf\beta}}- {\hat \imath^{*\alf\beta}} $ presented in
(\ref{i-i}) one can see that ${\hat \imath^{\alf\beta}_{DKO}} =
{\hat \imath^{\alf\beta}} $, where ${\hat \imath^{\alf\beta}} $ is
just the superpotential of the type constructed in
subsect.~\ref{sec:31}. However, recall that an additional divergence
in the Lagrangian in \cite{DerKatzOgushi} differs from the
divergence here defined by (\ref{divd}) in Appendix \ref{Appendix2},
see remarks around (\ref{divd}).

The result of the application here is that all the superpotentials
of the family give the same standard accepted mass for the S-AdS BH.
Thus, differences between various conserved quantities of the
family, possibly, look as not essential. However, numerus solutions
of popular gravitational theories frequently have very exotic
properties. Therefore, wider possibilities to study such solutions
are desirable, and, in this relation, the suggested family presents
a more universal instrument. We do not exclude the situation when
any solution any modified theory of gravity could be a crucial test
solution for a choice between members (conserved quantities) of the
family. We plan such applications in future.

\appendix

\section{Auxiliary algebraic expressions}
\m{Appendix1} 

Here, we give useful for calculations algebraic properties of the
operator presented by the notation ${\l. Q^A \r|}^\alf_\beta$
included in (\ref{LieQ}):
 \be
 \delta Q^A= {\pounds}_\xi Q^A =-\xi^\alf \di_\alf Q^A +
 {\l. Q^A \r|}^\alf_\beta \di_\alf \xi^\beta\, .
\m{LieQ+}
 \ee
In general, we follow to \cite{Mitzk}, however our treating ${\l.
Q^A \r|}^\alf_\beta$ is more simple and more effective, as we
imagine. We define the operator for covariant quantities only:
tensor densities or sets of tensor densities.  For example, for a
tensor density of the weight +$n$ one has
 \bea
\l. Q^{\alf\beta\ldots\gamma}_{\pi\rho\ldots\sig} \r|^\mu_\nu &=&
-n\delta^\mu_\nu Q^{\alf\beta\ldots\gamma}_{\pi\rho\ldots\sig}
+\delta^\alf_\nu
Q^{\mu\beta\ldots\gamma}_{\pi\rho\ldots\sig}+\delta^\beta_\nu
Q^{\alf\mu\ldots\gamma}_{\pi\rho\ldots\sig} +\ldots
+\delta^\gamma_\nu
Q^{\alf\beta\ldots\mu}_{\pi\rho\ldots\sig}\nonumber\\&-&
\delta^\mu_\pi Q^{\alf\beta\ldots\gamma}_{\nu\rho\ldots\sig}-
\delta^\mu_\rho Q^{\alf\beta\ldots\gamma}_{\pi\nu\ldots\sig}-\ldots-
\delta^\mu_\sig Q^{\alf\beta\ldots\gamma}_{\pi\rho\ldots\nu}\,.
 \m{Q-tensor}
 \eea
Thus, for the metric one has $\l. g_{\mu\nu} \r|^\alf_\beta=
-\delta^\alf_\mu g_{\beta\nu}-\delta^\alf_\nu g_{\mu\beta}$, or for
the scalar density: $\l. \lag
\r|^\alf_\beta=-\delta^\alf_\beta\lag$. One can see that
calculations with expressions, like (\ref{Q-tensor}), could be very
cumbersome. Whereas the use of the abstract form ${\l. Q^A
\r|}^\alf_\beta$ is, indeed, more economical, this is a main reason
why we suggest to apply it here. Only to show a final result after
calculations the form (\ref{Q-tensor}) could be represented. Below
we describe properties of the definition $ \l. Q^A \r|^\alf_\beta$,
which are necessary in the present paper.

The right hand side of the equation (\ref{Q-tensor}) shows that the
quantities ${\l. Q^A \r|}^\alf_\beta$ are tensor densities of the
same weight as the tensor densities $Q^A$. Therefore, the covariant
derivative $\l({\l. Q^A \r|}^\alf_\beta\r)_{;\gamma}$ of tensor
densities ${\l. Q^A \r|}^\alf_\beta$ is defined in a usual manner as
applied to a covariant quantity. Thus, the evident property
 \be
\l. \l(Q^A{}_{;\alf}\r)\r|^\tau_\rho =\l.
\l(Q^A{}\r|^\tau_\rho\r)_{;\alf} - \delta^\tau_\alf Q^A{}_{;\rho}
 \m{line-deriv}
 \ee
follows after covariant differentiation of (\ref{Q-tensor}). Also an
action of  a double vertical line is defined by a natural way:
 \be
\l.\l.Q^A\r|^\alf_\beta\r|^\mu_\nu \equiv
\l.\l(\l.Q^A\r|^\alf_\beta\r)\r|^\mu_\nu \equiv{\l. Q^B \r|}^\mu_\nu
 \m{DoubleLine}
 \ee
where `${}^B$' is a new generalized index. More important properties
follow from the usual properties of the Lie derivative. Thus from
 \bea
 {\pounds}_\xi\delta^\rho_\tau &=&0\,;\m{LieDelta}\\
{\pounds}_\xi (Q^AP^B) &=& P^B{\pounds}_\xi (Q^A) + Q^A{\pounds}_\xi
(P^B)\,;\m{LiePQ}\\
{\pounds}_\zeta {\pounds}_\xi(Q^A) -
{\pounds}_\xi{\pounds}_\zeta(Q^A)& =& {\pounds}_{[\zeta\xi]}
(Q^A)\,, \m{LieAntisym}
 \eea
where
$[\zeta\xi]=\xi^\rho\zeta^\alf{}_{,\rho}-\zeta^\rho\xi^\alf{}_{,\rho}$,
the next relations are derived:
 \bea \l.\delta^\rho_\tau \r|^\alf_\beta &=&
0\,;
 \m{LieDelta+}\\
\l.(Q^AP^B)\r|^\alf_\beta &=& \l.(P^B)\r|^\alf_\beta Q^A
+\l.(Q^A)\r|^\alf_\beta P^B\,;
 \m{LiePQ+}\\
 \l. \l.Q^A \r|^\beta_\rho \r|^\tau_\alf -
 \l. \l.Q^A \r|^\tau_\alf \r|^\beta_\rho &=& \delta^\beta_\alf\l.Q^A\r|^\tau_\rho
-\delta^\tau_\rho\l.Q^A\r|^\beta_\alf\,.
 \m{LieAntisym+}
 \eea
Among these the second property (\ref{LiePQ+}) is more useful. Thus,
in our calculations frequently we use the transformations, like this
 \be
 \frac{\di \lag}{\di Q_{B;\alf}}\l.Q_B\r|^\rho_\tau =
\l. \l(\frac{\di \lag}{\di Q_{B;\alf}}Q_B\r)\r|^\rho_\tau -\l.
\l(\frac{\di \lag}{\di Q_{B;\alf}}\r)\r|^\rho_\tau Q_B\,.
 \m{(LagQ)Q}
 \ee
The first term at the right hand side, to which the vertical line is
applied, is a vector density, therefore it is useful also
   \be
\l. \l(\frac{\di \lag}{\di Q_{B;\alf}}Q_B\r)\r|^\rho_\tau =  -
\delta^\rho_\tau\frac{\di \lag}{\di Q_{B;\alf}}Q_B +
\delta^\alf_\tau\frac{\di \lag}{\di Q_{B;\rho}}Q_B\,.
 \m{(LagQ)Q+}
 \ee
The above properties are enough to derive the useful relation:
 \be
\frac{\di \l(\l.Q^A\r|^\alf_\beta\r)}{\di Q^B}\l.Q^B\r|^\rho_\tau =
\l.\l.Q^A\r|^\alf_\beta\r|^\rho_\tau\,. \m{ACC}
 \ee

The definition $ \l. Q^A \r|^\alf_\beta$ in (\ref{LieQ+})
corresponds to the presentation of the covariant derivative by the
way:
 \be
\Bar D_\alf Q^A = \di_\alf Q^A +
 {\l. Q^A \r|}^\rho_\tau \Bar \Gamma_{\alf\rho}^\tau \, .
\m{LieQ+A}
 \ee
Namely this presentation is used to represent a partial derivative
through a covariant one. Recall that for the vector density $ \hat
Q^\alf$: $\Bar D_\alf \hat Q^\alf = \di_\alf \hat Q^\alf$, and for
the antisymmetric tensor density $\hat Q^{\alf\beta}$: $\Bar D_\beta
\hat Q^{\alf\beta} = \di_\beta \hat Q^{\alf\beta}$. Then, the
definition (\ref{LieQ+A}) evidently gives
 \bea
{\l. \hat Q^\alf \r|}^\rho_\tau \Bar \Gamma_{\alf\rho}^\tau &=& 0\,,
 \m{d-vector}\\
{\l. \hat Q^{\alf\beta} \r|}^\rho_\tau \Bar \Gamma_{\beta\rho}^\tau
&=& 0\,. \m{d-tensor}
 \eea
Then (\ref{(LagQ)Q+}), as an example of the vector density,  gives:
 \be
 \l. \l(\frac{\di
\lag}{\di Q_{B;\alf}}Q_B\r)\r|^\rho_\tau\Bar \Gamma_{\alf\rho}^\tau
=0\,.
 \m{2example}
 \ee
At last, we note that the notation $ \l. Q^A \r|^\alf_\beta$  can be
used also in antisymmetrization of the covariant derivatives:
 \be
 Q^A{}_{;\mu\nu}-Q^A{}_{;\nu\mu} =\l. Q^A \r|^\alf_\beta \Bar
 R_\alf{}^\beta{}_{\mu\nu}\,.
 \m{line-Riemann}
 \ee

Calculations in the text of the paper are very prolonged and it is
impossible to note in each the case what formulae in this Appendix
have been used. Therefore, we do not do it, explaining only a
general direction of calculations. The information in this Appendix
is quite enough to repeat our calculations without principal
obstacles.

\section{Necessary formulae in the
Einstein-Gauss-Bonnet gravity}
 \m{Appendix2}

The action of the Einstein $D$-dimensional theory with a bare
cosmological term $\Lambda_0$ corrected by the Gauss-Bonnet term
(see, for example, \cite{DT2}) is
 \bea
 S  &=& -\frac{1}{2\k_D}\int d^D x\hat L_{EGB} +\int d^D x\lag_{m}
\nonumber\\&=& -\frac{1}{2\k_D}\int d^D x \sqrt{-g} \l[R -
2\Lambda_0 +
 \alpha(RR)_{GB}\r]  +\int d^D x\lag_{m}\,,
 \m{EGBaction}
 \eea
 \be
(RR)_{GB} \equiv  R_{\alf\beta\gamma\delta}
R^{\alf\beta\gamma\delta}
 - 4 R_{\alf\beta} R^{\alf\beta}+ R^2 \,,
 \m{GB-term}
 \ee
where $\k_D = 2\Omega_{D-2}G_D> 0$  and $\alpha >0$; $G_D$ is the
$D$-dimension Newton's constant, $\Omega_{D-2}$ is the area of a
unit $(D-2)$-dimensional sphere, and we restrict ourselves by
$\Lambda_0 \leq 0$. The subscript `${}_{E}$' is related to the pure
Einstein part of the action (\ref{EGBaction}), and the subscript
`${}_{GB}$' is related to the Gauss-Bonnet part connected with
$\alf$-coefficient.

To present the metric Lagrangian $\hat L_{EGB}$ in an explicitly
covariant form $\lag_{EGB}$ one has to change partial derivatives
$\di_\mu$ of the dynamic metric $g_{\mu\nu}$  in the Riemannian
tensor by the covariant derivatives $\Bar D_\mu$. We use next useful
formulae:
 \be
 R^\lam{}_{\tau\rho\sig} =
\BD_\rho \Delta^\lam_{\tau\sig} -  \BD_\sig\Delta^\lam_{\tau\rho} +
 \Delta^\lam_{\rho\eta} \Delta^\eta_{\tau\sig} -
 \Delta^\lam_{\eta\sig} \Delta^\eta_{\tau\rho}
 + \Bar R^\lam{}_{\tau\rho\sig} = \delta R^\lam{}_{\tau\rho\sig} +
 \Bar R^\lam{}_{\tau\rho\sig}\,
 \m{(17-DmD)}
 \ee
 where
 \be
\Del^\alpha_{\mu\nu} = \Gamma^\alpha_{\mu\nu} - \Bar
{\Gamma}^\alpha_{\mu\nu} = \half g^{\alf\rho}\l( \BD_\mu g_{\rho\nu}
+ \BD_\nu g_{\rho\mu} - \BD_\rho g_{\mu\nu}\r)\, \m{DeltaDefD}
  \ee
is the difference between the Christoffel symbols related to the
dynamic $g_{\mu\nu}$ and the background $\Bar g_{\mu\nu}$ metrics.
It is useful the next relations also. Analogously to (\ref{LieQ+A})
one defines the covariant derivative related to the dynamic metric:
 \be
D_\alf Q^A = \di_\alf Q^A +
 {\l. Q^A \r|}^\rho_\tau \Gamma_{\alf\rho}^\tau \, .
\m{LieQ+Ad}
 \ee
Then, comparing (\ref{LieQ+Ad}) and (\ref{LieQ+A}) one obtains
 \be
D_\alf Q^A = \Bar D_\alf Q^A +
 {\l. Q^A \r|}^\rho_\tau \Delta_{\alf\rho}^\tau \, .
\m{LieQ+Add}
 \ee

The coefficients ${n^*}$ and ${m^*}$ (see formulae (\ref{nEGB}) and
(\ref{mEGB}) below), corresponding to the Lagrangian $\lag_{EGB}$ in
(\ref{EGBaction}), are necessary for calculating superpotentials.
However, at the first it is useful to present the next derivatives:
 \bea
&{}& \frac{-2\k_D}{\sqrt{-g}}\frac{\di \lag_{EGB}}{\di
g_{\mu\nu;\alf}} =
 \frac{-2\k_D}{\sqrt{-g}}\l(\frac{\di \lag_{E}}{\di
g_{\mu\nu;\alf}}+
 \frac{\di \lag_{GB}}{\di
g_{\mu\nu;\alf}} \r)\nonumber\\& =& 2
\l[\Delta^{\alf}_{\sig\rho}g^{\sig[\rho}g^{\mu]\nu} +
g^{\alf\sig}\Delta^{(\mu}_{\sig\rho}g^{\nu)\rho} -
g^{\alf(\mu}\Delta^{\nu)}_{\sig\rho}g^{\sig\rho} \r]\nonumber\\& +&
4\alf\l[2R^{\alf\sig\rho(\mu}\Delta^{\nu)}_{\sig\rho}
-\Delta^{\alf}_{\sig\rho}R^{\sig\mu\nu\rho}\r]\nonumber\\& -&
4\alf\l[2 R^{\alf\sig}\Delta^{(\mu}_{\sig\rho}g^{\nu)\rho} -
2g^{\alf(\mu}\Delta^{\nu)}_{\sig\rho}R^{\sig\rho} + 2
g^{\alf\sig}\Delta^{(\mu}_{\sig\rho}R^{\nu)\rho}
-2R^{\alf(\mu}\Delta^{\nu)}_{\sig\rho}g^{\sig\rho}\r.\nonumber\\ &+&
\l. \Delta^{\alf}_{\sig\rho}R^{\sig\rho}g^{\mu\nu}+
\Delta^{\alf}_{\sig\rho}g^{\sig\rho}R^{\mu\nu} -
2\Delta^{\alf}_{\sig\rho}R^{\sig(\mu}g^{\nu)\rho}\r]
 \nonumber\\& +&
 4\alf R
\l[\Delta^{\alf}_{\sig\rho}g^{\sig[\rho}g^{\mu]\nu} +
g^{\alf\sig}\Delta^{(\mu}_{\sig\rho}g^{\nu)\rho} -
g^{\alf(\mu}\Delta^{\nu)}_{\sig\rho}g^{\sig\rho} \r]\,;
 \m{DiLDgmn}
 \eea
 \bea
&{}& \frac{-2\k_D}{\sqrt{-g}}\frac{\di \lag_{EGB}}{\di
g_{\mu\nu;\alf\beta}} =
 \frac{-2\k_D}{\sqrt{-g}}\l(\frac{\di \lag_{E}}{\di
g_{\mu\nu;\alf\beta}}+
 \frac{\di \lag_{GB}}{\di
g_{\mu\nu;\alf\beta}} \r)\nonumber\\& =&
\l[g^{\alf(\mu}g^{\nu)\beta}- g^{\alf\beta}g^{\mu\nu}\r]\nonumber\\&
+& 2\alf\l[2R^{\alf(\mu\nu)\beta} - 4 R^{\alf(\mu}g^{\nu)\beta} +
2g^{\mu\nu}R^{\alf\beta} + 2g^{\alf\beta}R^{\mu\nu} + R
\l(g^{\alf(\mu}g^{\nu)\beta}- g^{\alf\beta}g^{\mu\nu} \r)\r]\,.
  \m{DiLDDgmn}
 \eea
Substituting (\ref{DiLDgmn}) and (\ref{DiLDDgmn}) into
(\ref{n-inverse}) and (\ref{m-inverse}) one obtains
 \bea
 {\hat n^*}_{\sig}{}^{\lam\alf\beta} &= &{}_{(E)}{\hat
 n^*}_{\sig}{}^{\lam\alf\beta} + {}_{(GB)}{\hat n^*}_{\sig}{}^{\lam\alf\beta}
\m{nEGB} \\ &=& \frac{\sqrt{-
g}}{2\k_D}\l\{g^{\alf\beta}\delta^\lam_\sig
-g^{\lam(\alf}\delta^{\beta)}_\sig \r\}
 \nonumber\\ &+& \frac{\alf\sqrt{-g}}{\k_D}
 \l\{-2R_\sig{}^{(\alf\beta)\lam} - 4 R_\sig^{\lam}g^{\alf\beta} +
 4 R_\sig^{(\alf}g^{\beta)\lam} + R\l(g^{\alf\beta}\delta^\lam_\sig
-g^{\lam(\alf}\delta^{\beta)}_\sig\r)\r\}\, .
 \nonumber
 \eea
 \bea
 {\hat m^*}_{\sig}{}^{\alf\beta} &= & {}_{(E)}{\hat m^*}_{\sig}{}^{\alf\beta}
 +{}_{(GB)}{\hat m^*}_{\sig}{}^{\alf\beta}
 \m{mEGB}\\ &=&
-\frac{\sqrt{- g}}{2\k_D}\l[\delta^\alf_\sig
\Delta^{\beta}_{\rho\tau}g^{\rho\tau} - 2
\Delta^{\alf}_{\sig\rho}g^{\beta\rho} + \Delta^{\rho}_{\rho\sig}
g^{\alf\beta}  \r]
 \nonumber\\ &+ & \frac{2\alf\sqrt{-g}}{\k_D}
 \l[\ R^{\alf\tau\rho}{}_\sig \Delta^{\beta}_{\tau\rho}
-2R^{\alf(\tau\beta)}{}_\rho\Delta^{\rho}_{\tau\sig}
\r] \nonumber\\
&+ & \frac{4\alf\sqrt{-g}}{\k_D}\l[4 g^{\rho[\alf}
R^{\beta]}_{\tau}\Delta^\tau_{\rho\sig} + 2 R^{[\alf}_\sig
g^{\tau]\rho}\Delta^{\beta}_{\tau\rho}+ 2 g^{\alf[\beta}
R^{\tau]}_{\rho}\Delta^{\rho}_{\tau\sig} \r.
 \nonumber\\&-& \l. g^{\tau\beta}\l(\BD_{(\tau} R^\alf_{\sig)}
  + R^{\rho}_{(\tau}
\Delta^{\alf}_{\sig)\rho} - R^{\alf}_\rho
\Delta^{\rho}_{\tau\sig}\r) \r]
 \nonumber\\&-&
 \frac{\alf\sqrt{- g}}{\k_D}\l[\l(\delta^\alf_\sig
\Delta^{\beta}_{\rho\tau}g^{\rho\tau} - 2
\Delta^{\alf}_{\sig\rho}g^{\beta\rho} + \Delta^{\rho}_{\rho\sig}
g^{\alf\beta}\r)R - 2\delta^{(\alf}_\sig g^{\tau)\beta} \di_\tau R
\r]\,.\nonumber
 \eea
Using the coefficients (\ref{nEGB}) and (\ref{mEGB}) in
(\ref{(+11+star)}) one obtains for a pure N{\oe}ther canonical
starred superpotential in EGB gravity
  \bea
\hat \imath^{*\alf\beta} &= & {1\over \k_D}\big({\hat
g^{\rho[\alf}\Bar D_{\rho} \xi^{\beta]}} + \hat
g^{\rho[\alf}\Delta^{\beta]}_{\rho\sig}\xi^\sig\big)
 \nonumber\\ &- & \frac{2\alf\sqrt{-g}}{\k_D}
 \l\{\Delta^{\rho}_{\lam\sig}R_\rho{}^{\lam\alf\beta} +
4\Delta^{\rho}_{\lam\sig}g^{\lam[\alf}R^{\beta]}_\rho +
\Delta^{[\alf}_{\rho\sig}g^{\beta]\rho}R\r\}\xi^\sig
 \nonumber\\&-&
\frac{2\alf\sqrt{-g}}{\k_D}\l\{R_\sig{}^{\lam\alf\beta} +
 4 g^{\lam[\alf}R^{\beta]}_\sig + \delta_\sig^{[\alf}g^{\beta]\lam}R
 \r\}\BD_\lam \xi^\sig\, ,
 \m{CanSupEGB}\\
 \Bar{\hat \imath^{*\alf\beta}}  &= &
{1\over \k_D} \Bar {D^{[\alf} \hat\xi^{\beta]}}
 -
\frac{2\alf}{\k_D}\l\{\Bar{\hat R}_\sig{}^{\lam\alf\beta} +
 4 \Bar{g^{\lam[\alf}\hat R^{\beta]}_\sig} + \delta_\sig^{[\alf}
 \Bar{g^{\beta]\lam}\hat R}
 \r\}\BD_\lam \xi^\sig\, .
 \m{BarCanSupEGB}
 \eea
To finalize forming a pure canonical superpotential one needs to
choose the divergence. We prefer the choice induced by the
Katz-Lifshits approach \cite{KatzLivshits} instead of the choice in
\cite{DerKatzOgushi} (see discussions in
\cite{Petrov2009,Petrov2009cor,Petrov2009a}). Thus, we choose in
(\ref{7A'}) - (\ref{7A'''})
 \be
\hat d^\alf =  {}_{(E)}\hat d^\alf + {}_{(GB)}\hat d^\alf = {2}
\Delta^{[\tau}_{\tau\beta}\hat g^{\alf]\beta}  +
 4 \alf\l(
\hat R_\rho{}^{\beta\tau\alf} - 2\hat R^{[\tau}_\rho g^{\alf]\beta}-
2\delta^{[\tau}_\rho \hat R^{\alf]\beta}+ \delta^{[\tau}_\rho
g^{\alf]\beta} \hat R\r)\Delta^\rho_{\tau\beta}\, .
 \m{divd}
 \ee

Using the coefficients (\ref{nEGB}) and (\ref{mEGB}) in
(\ref{(+16+star)}) one obtains the Belinfante corrected starred
superpotential in EGB gravity
  \bea
\hat \imath^{*\alf\beta}_B  &= & {1\over \k_D}\l[\l(
\delta^{[\alf}_\sig \BD_\lam \hat g^{\beta]\lam} -\BD^{[\alf}\hat
g^{\beta]\rho}\Bar g_{\rho\sig} \r)\xi^\sig +\hat
g^{\lam[\alf}\BD_\lam \xi^{\beta]}\r]\nonumber\\
&+&{\alf\over \k_D}\BD_\lam\l\{\hat R_\sig{}^{\lam\alf\beta} +
 4 g^{\lam[\alf}\hat R^{\beta]}_\sig
 +\l[2\hat R_\tau{}^{\rho\lam[\alf}
-2\hat R^{\rho\lam}{}_\tau{}^{[\alf} - 8\hat R^\lam_\tau
g^{\rho[\alf}\r.\r.
 \nonumber\\&+& \l.\l.4
\hat R^\rho_\tau g^{\lam[\alf} +4 g^{\rho\lam} \hat R^{[\alf}_\tau +
2\hat R\l( \delta^\lam_\tau g^{\rho[\alf}- \delta^\rho_\tau
g^{\lam[\alf}\r)\r]\Bar g^{\beta]\tau}\Bar g_{\rho\sig} \r\}\xi^\sig
 \nonumber\\&-&{2\alf\over \k_D}\l\{{\hat R}_\sig{}^{\lam\alf\beta} +
 4 {g^{\lam[\alf}\hat R^{\beta]}_\sig} + \delta_\sig^{[\alf}
 g^{\beta]\lam}\hat R
 \r\}
 \BD_\lam \xi^\sig\,,
 \m{BelSupEGB}\\
\Bar{\hat \imath^{*\alf\beta}_{B}}  &=& {1\over \k_D}\Bar{\hat
g^{\lam[\alf}D_\lam \xi^{\beta]}}-{2\alf\over \k_D}\l\{\Bar{\hat
R}_\sig{}^{\lam\alf\beta} +
 4 \Bar{g^{\lam[\alf}\hat R^{\beta]}_\sig} + \delta_\sig^{[\alf}
 \Bar{g^{\beta]\lam}\hat R}
 \r\}
 \BD_\lam \xi^\sig\,.
 \m{BarBelSupEGB}
 \eea

%
%


\bibliographystyle{spphys}       
\bibliography{Petrov_Lompay}   

\begin{thebibliography}{10}
\providecommand{\url}[1]{{#1}}
\providecommand{\urlprefix}{URL }
\expandafter\ifx\csname urlstyle\endcsname\relax
  \providecommand{\doi}[1]{DOI \discretionary{}{}{}#1}\else
  \providecommand{\doi}{DOI \discretionary{}{}{}\begingroup
  \urlstyle{rm}\Url}\fi

\bibitem{Petrov2008}
A.N. Petrov, in \emph{Classical and Quantum Gravity}, ed. by M.N. Christiansen,
  T.K. Rasmussen (Nova Science Publishers, New York, 2008), chap.~2, pp.
  79--160.
\newblock Preprint arXiv:0705.0019 [gr-qc]

\bibitem{Szabados}
L.B. Szabados, Living Rev. Relativity \textbf{12}(4), 4 (2009).
\newblock Online version: http://www.livingreviews.org/lrr-2009-4

\bibitem{PittsSchive2001b}
J.B. Pitts, W.C. Schieve.
\newblock Null cones in lorentz-covariant general relativity (2001).
\newblock Preprint arXiv:gr-qc/0111004

\bibitem{Bergmann58}
P.G. Bergmann, Phys. Rev. \textbf{112}(1), 287 (1958)

\bibitem{Komar59}
A.~Komar, Phys. Rev. \textbf{113}(3), 934 (1959)

\bibitem{Trautman62}
A.~Trautman, in \emph{Gravitation: an Introduction to Current Research}, ed. by
  L.~Witten (John Wiley and Sons, New York - London, 1962), chap.~5, pp.
  169--198

\bibitem{[11]}
S.~Deser, Gen. Relat. Grav. \textbf{1}(1), 9 (1970).
\newblock Preprint gr-qc/0411023

\bibitem{GPP}
L.P. Grishchuk, A.N. Petrov, A.D. Popova, Commun. Math. Phys. \textbf{94}(3),
  379 (1984)

\bibitem{KBL}
J.~Katz, J.~Bi\v{c}\'{a}k, D.~Lynden-Bell, Phys. Rev. D \textbf{55}(10), 5957
  (1997).
\newblock Preprint arXiv:gr-qc/0504041

\bibitem{ChenNester}
C.M. Chen, J.M. Nester, Class. Quantum Grav. \textbf{16}(4), 1279 (1999).
\newblock Preprint arXiv:gr-qc/9809020

\bibitem{Nester10}
C.M. Chen, J.M. Nester, Grav. Cosmol. \textbf{6}(4), 257 (2000).
\newblock Preprint arXiv:gr-qc/0001088

\bibitem{Fatibene-etal}
L.~Fatibene, M.~Ferraris, M.~Francaviglia, M.~Raiteri, J. Math. Phys.
  \textbf{42}(3), 1173 (23pp) (2001).
\newblock Preprint arXiv:gr-qc/0003019

\bibitem{PK}
A.N. Petrov, J.~Katz, Proc. R. Soc. A, London \textbf{458}(2018), 319 (2002).
\newblock Preprint arXiv:gr-qc/9911025

\bibitem{DT2}
S.~Deser, B.~Tekin, Phys. Rev. D \textbf{67}(8), 084009 (7pp) (2003).
\newblock Preprint arXiv:hep-th/0212292

\bibitem{Lovelock}
D.~Lovelock, J. Math. Phys. \textbf{12}(3), 498 (4pp) (1971)

\bibitem{Sotiriou}
T.P. Sotiriou, V.~Faraoni, Rev. Mod. Phys. \textbf{82}(1), 451 (2010)

\bibitem{DerKatzOgushi}
N.~Deruelle, J.~Katz, S.~Ogushi, Class. Quantum Grav. \textbf{21}(8), 1971
  (2004).
\newblock Preprint arXiv:gr-qc/0310098

\bibitem{Fatibene1}
L.~Fatibene, M.~Ferraris, M.~Francaviglia, Int. J. Geom. Meth. Mod. Phys.
  \textbf{2}(3), 373 (2005).
\newblock Preprint arXiv:math-ph/0411029

\bibitem{KatzLivshits}
J.~Katz, G.I. Livshits, Class. Quantum Grav. \textbf{25}(17), 175024 (17pp)
  (2008).
\newblock Preprint arXiv:0807.3079 [gr-qc]

\bibitem{Petrov2009}
A.N. Petrov, Class. Quantum Grav. \textbf{26}(7), 135010 (16pp) (2009).
\newblock Corrigendum in: \cite{Petrov2009cor}; Preprint arXiv:0905.3622
  [gr-qc]

\bibitem{Petrov2009cor}
A.N. Petrov, Class. Quantum Grav. \textbf{27}(6), 069801 (2pp) (2010).
\newblock Corrigendum to the ref. \cite{Petrov2009}

\bibitem{Petrov2009a}
A.N. Petrov, Grav. Cosmol. \textbf{16}(1), 34 (2010).
\newblock Preprint arXiv:0911.5419 [gr-qc]

\bibitem{Petrov2011}
A.N. Petrov, Class. Quantum Grav. \textbf{28}(7), 215021 (17pp) (2011).
\newblock Preprint arXiv:1102.5636 [gr-qc]

\bibitem{Mitzk}
N.V. Mitzkevich, \emph{Physical Fields in General Theory of Relativity} (Nauka,
  Moscow, 1969).
\newblock In Russian

\bibitem{Ray}
J.R. Ray, Nuovo Cim. A \textbf{56}(1), 189 (1968)

\bibitem{Belinfante}
F.J. Belinfante, Physica \textbf{6}(9), 887 (1939)

\bibitem{BD+}
D.G. Boulware, S.~Deser, Phys. Rev. Lett. \textbf{55}(24), 2656 (1985)

\bibitem{Klein18}
F.~Klein, Nachr. d. Konig. Gesellsch. d. Wiss. zu Gottingen, Math-phys. Klasse
  pp. 171--189 (1918).
\newblock Reprinted in: \cite{Klein21}, pages 568-585

\bibitem{Klein21}
F.~Klein, \emph{Gesammelte Mathematische Abhandlungen. Band 1} (Springer,
  Berlin, 1921)

\bibitem{Kosmann-Schwarzbach}
Y.~Kosmann-Schwarzbach, \emph{The Noether Theorems}.
\newblock Sources and studies in the history of mathematics and physics
  (Springer, New York - Dordrecht - Heidelberg - London, 2011).
\newblock Translated by Bertram E. Schwarzbach

\bibitem{KonoplevaPopov}
N.P. Konopleva, V.N. Popov, \emph{Gauge Fields} (Atomizdat, Moscow, 1980).
\newblock In Russian

\bibitem{NinaByers}
N.~Byers.
\newblock E. noether's discovery of the deep connection between symmetries and
  conservation laws (1998).
\newblock Preprint arXiv:physics/9807044 [physics.hist-ph]; (21pp)

\bibitem{BradingBrown}
K.~Brading, H.R. Brown.
\newblock Noether's theorems and gauge symmetries (2000).
\newblock Preprint arXiv:hep-th/0009058; (16pp)

\bibitem{BradingCastellani}
K.~Brading, H.R. Brown, in \emph{Symmetries in Physics. Philosophical
  reflections}, ed. by K.~Brading, E.~Castellani (CUP, Cambridge, 2003), pp.
  89--109

\bibitem{Sardanashvily}
G.~Sardanashvily, Int. J. Geom. Methods Mod. Phys. \textbf{6}(6), 1047 (2009).
\newblock Preprint arXiv:0906.1732 [math-ph]

\bibitem{Noether1918}
E.~Noether, Nachr. d. Konig. Gesellsch. d. Wiss. zu Gottingen, Math-phys.
  Klasse pp. 235--257 (1918).
\newblock In German. English translation in: \cite{Kosmann-Schwarzbach}, pages
  3-19

\bibitem{Freud39}
P.~Von~Freud, Ann. of Math. \textbf{40}(2), 417 (1939)

\bibitem{Papapetrou48}
A.~Papapetrou, Proc. R. Irish Acad. A \textbf{52}(2), 11 (1948)

\bibitem{Paddila}
A.~Paddila, Class. Quantum Grav. \textbf{20}(14), 3129 (2003).
\newblock Preprint arXiv:gr-qc/0303082

\bibitem{AFrancavigliaR}
G.~Allemandi, M.~Francaviglia, M.~Raiteri, Class. Quantum Grav.
  \textbf{20}(23), 5103 (2003).
\newblock Preprint arXiv:gr-qc/0308019

\bibitem{Okuyama}
N.~Okuyama, J.I. Koga, Phys. Rev. D \textbf{71}(8), 084009 (9pp) (2005).
\newblock Preprint arXiv:hep-th/0501044

\end{thebibliography}

%
%

\end{document}